\documentclass[twocolumn, tighten]{aastex62}

\usepackage{color}
\usepackage{amsmath}
\usepackage{natbib}
\hypersetup{breaklinks=true, colorlinks=true, urlcolor={blue}, citecolor={blue}, linkcolor={blue}}

\usepackage{etoolbox}
\shortauthors{Hu et al.}
\accepted{Sep.~27, 2019}

\makeatletter
\patchcmd{\NAT@citex}
  {\@citea\NAT@hyper@{%
     \NAT@nmfmt{\NAT@nm}%
     \hyper@natlinkbreak{\NAT@aysep\NAT@spacechar}{\@citeb\@extra@b@citeb}%
     \NAT@date}}
  {\@citea\NAT@nmfmt{\NAT@nm}%
   \NAT@aysep\NAT@spacechar\NAT@hyper@{\NAT@date}}{}{}

\patchcmd{\NAT@citex}
  {\@citea\NAT@hyper@{%
     \NAT@nmfmt{\NAT@nm}%
     \hyper@natlinkbreak{\NAT@spacechar\NAT@@open\if*#1*\else#1\NAT@spacechar\fi}%
       {\@citeb\@extra@b@citeb}%
     \NAT@date}}
  {\@citea\NAT@nmfmt{\NAT@nm}%
   \NAT@spacechar\NAT@@open\if*#1*\else#1\NAT@spacechar\fi\NAT@hyper@{\NAT@date}}
  {}{}

\makeatother

\newcommand{\chandra}{\emph{Chandra}}

\newcommand{\xmm}{\emph{XMM-Newton}}
\newcommand{\swift}{\emph{Swift}}
\newcommand{\rxte}{\emph{RXTE}}

\newcommand{\fermi}{\emph{Fermi}}

\newcommand{\nustar}{\emph{NuSTAR}}

\newcommand{\lumcgs}{erg~s$^{-1}$}

\begin{document}
\title{Monitoring the Superorbital Period Variation and the Spin Period Evolution of SMC X-1}
\author{Chin-Ping Hu}
\altaffiliation{JSPS International Research Fellow}
\affiliation{Department of Astronomy, Kyoto University, Kitashirakawa-Oiwake-cho, Sakyo-ku, Kyoto 606-8502, Japan}
\author{Tatehiro Mihara}
\affiliation{MAXI team, Institute of Physical and Chemical Research (RIKEN), 2-1, Hirosawa, Wako, Saitama 351-0198, Japan}
\author{Mutsumi Sugizaki}
\affiliation{Department of Physics, Tokyo Institute of Technology, 2-12-1 Ookayama, Meguro-ku, Tokyo 152-8551, Japan}
\author{Yoshihiro Ueda}
\affiliation{Department of Astronomy, Kyoto University, Kitashirakawa-Oiwake-cho, Sakyo-ku, Kyoto 606-8502, Japan}
\author{Teruaki Enoto}
\affiliation{Department of Astronomy, Kyoto University, Kitashirakawa-Oiwake-cho, Sakyo-ku, Kyoto 606-8502, Japan}
\affiliation{The Hakubi Center for Advanced Research, Kyoto University, Kitashirakawa-Oiwake-cho, Sakyo-ku, Kyoto 606-8502, Japan}

\correspondingauthor{C.-P. Hu}
\email{cphu@kusastro.kyoto-u.ac.jp}

\begin{abstract}
The X-ray pulsar SMC X-1 shows a superorbital modulation with an unstable cycle length in the X-ray band. We present its timing behaviors, including the spin, orbital, and superorbital modulations, beyond the end of the \emph{Rossi X-ray Timing Explorer} mission. The time-frequency maps derived by the wavelet $Z$-transform and the Hilbert-Huang transform suggest that a new superorbital period excursion event occurred in $\sim$MJD 57100 (2015 March). This indicates the excursion is recurrent and probably (quasi)periodic. The hardness ratio obtained with the Monitor of All-sky X-ray Image (MAXI) suggests increased absorption during the transition from the high to the low state in the superorbital cycle. Compared to the regular epochs, the superorbital profile during the excursion epochs has a shallower and narrower valley, likely caused by a flatter warp. By tracking the spin period evolution with the MAXI Gas Slit Camera in 2--20\,keV, we derive an averaged spin-up rate of $\dot{\nu}=2.515(3)\times10^{-11}$\,s$^{-2}$ during the period between MJD 55141 (2009 November) and 58526 (2019 February). We obtain no positive correlation between the spin frequency residual and the superorbital frequency, but a torque change accompanying the superorbital period excursion is possible. We suggest that the accretion torque on the neutron star could be changed by various mechanisms, including the change of mass accretion rate and the warp angle. We update the value of the orbital decay as $\dot{P}_{\rm{orb}}/P_{\rm{orb}}=-3.380(6)\times10^{-6}$\,yr$^{-1}$. Finally, we reconfirm the detection of the superorbital modulation in the optical band and its coherence in phase with the X-ray modulation. 
\end{abstract}


\section{Introduction}\label{introduction}
Accreting pulsars in X-ray binaries are among the brightest and most variable X-ray emitters in the sky. The evolution of their (quasi)periodic signals, including the spin, orbital, and superorbital modulations, strongly depend on the accretion mechanisms \citep{Corbet1986, CorbetK2013}. Tracking their timing behaviors is helpful for understanding the connection between their accretion torque, the variability of the mass accretion rate, and the change of disk/accretion configuration. For example, the spin-up rates of OAO 1657$-$415 and 4U 0114+650 are different from their spin-down rates, implying the formation of a transient disk \citep{JenkeFW2012, HuCN2017}.  The superorbital modulation is usually interpreted by the precession of the radiation-driven warped disk model \citep{Pringle1996, OgilvieD2001}. The varying occultation of X-ray emission by the warped disk is thought to be the mechanism of the flux variability. The spin behavior could be associated with the change of the superorbital modulation. For instance, the pulsar in Her X-1 spins down during the anomalous low states, implying a change of disk warp angle or a change of mass accretion rate \citep{ParmarO1999, StaubertKP2009, Leahy2010}. Recently, it was proposed that the tidal force-induced precession of a ring on a tilted accretion disk \citep{Inoue2012, Inoue2019} could be an alternative origin. The precession of such a ring tube could pass through our line of sight to the central compact object quasiperiodically and cause the superorbital modulation.

The high-mass X-ray binary (HMXB) SMC X-1 consists of a 1.2 $M_{\odot}$ neutron star and an 18 $M_{\odot}$ supergiant companion \citep{Reynolds1993, vanderMeer2007, Falanga2015}. Its peak luminosity reaches $\gtrsim5\times10^{38}$\,\lumcgs, suggesting that SMC X-1 is a mildly super-Eddington source \citep{PriceGR1971, UlmerBW1973, CoeBE1981, LiV1997}. The neutron star is obscured by the companion every $3.89$\,days \citep{Schreier1972}. From the phase evolution of the mid-eclipse time before 2000, a decay of the orbital period is observed \citep{LevineRD1993, Wojdowski1998}. The value of the orbital period derivative is updated as $\dot{P}_{\rm{orb}}=-3.78(15)\times10^{-8}$\,day\,day$^{-1}$ \citep[or $\dot{\nu}_{\rm{orb}}=2.493(1)\times10^{-9}$\,day$^{-2}$,][]{Falanga2015}.

The long-term X-ray flux of SMC X-1 shows a high-low state transition with a timescale of $\sim60$ days \citep[e.g.,][]{Gruber1984}. Later long-term observation with the all-sky monitor (ASM) on board the \emph{Rossi X-ray Timing Explorer} (\rxte) suggests that the cycle length of this superorbital modulation is unstable \citep{Wojdowski1998, ClarksonCC2003a}. The time-frequency map obtained from the ASM light curve shows two excursion events in $\sim$MJD 50800 and $\sim$MJD 54000, where the superorbital period evolves to $P_{\rm{sup}}\sim40$\,days  \citep{Trowbridge2007, Hu2011}. Except for these two events, the superorbital period is relatively stable at $P_{\rm{sup}}\sim56$\,days, though short-term variabilities remain possible. 


The compact object in SMC X-1 is an X-ray pulsar with a spin period of $0.7$\,s \citep{Lucke1976}. The pulse profile varies with the energy bands, orbital phase, and superorbital phase. The hard X-ray pulse profile is relatively stable and superorbital phase-independent, suggesting that it is originated from the pencil beam from the polar cap \citep{Neilsen2004}. On the other hand, the soft X-ray profile is highly variable with superorbital phase \citep{Wojdowski1998, Neilsen2004}. It is interpreted as the reprocessing of X-rays from the accretion disk \citep{Hickox2005}. Since 1997, SMC X-1 has been well monitored with several X-ray instruments, and the most frequent monitoring was carried out with \rxte\ between 1996 and 2000 \citep{HenryS1977, Inam2010}. The full history of the pulse frequency evolution of SMC X-1 up to MJD 52987 (2003 December 14) is reported by \citet{Inam2010}. The frequency almost increases with time monotonically, but the spin-up rate changes significantly. The spin history can be divided into five epochs with a frequency derivative varying in the range of $(2$ -- $3.6)\times10^{-11}$\,s$^{-2}$. The spin-up rate could be associated with the drift of the superorbital modulation period \citep{DageCC2018}. Their correlation remains inconclusive through a detailed analysis, but the superorbital period excursion on $\sim$MJD 50800 coincides with a change in the spin-up rate.

The major goal of this research is to explore the connection between the spin, orbital, and superorbital modulations. With the ongoing observation made with the \emph{Neil Gehrels Swift Observatory} (hereafter \swift) and the Monitor of All-sky X-ray Image (MAXI), we extend $\sim$7 yr of the time baseline after the \rxte\ era. We would like to confirm whether the superorbital period excursion is recurrent and examine whether the short-term period drifting is real. We also process the new \chandra, \xmm, and \swift\ X-Ray Telescope (XRT) observations, as well as the photon events collected with MAXI, to extend the history of the pulse frequency measurements. This is used to examine whether there is any connection between the spin and the superorbital modulations. Moreover, we revisit the superorbital dependence of the orbital profile using MAXI data to see if the orbital profile in different superorbital phases is energy-dependent. Section \ref{observation} introduces the observation and data reduction. Then, we present the analysis result for the superorbital period excursion, orbital ephemeris and profile, and spin period evolution in Section \ref{result}. We also analyze the optical data taken with the All-Sky Automated Survey for Supernovae \citep[ASAS-SN; ][]{ShappeePG2014, KochanekSS2017} to examine the orbital and superorbital modulation in the optical band. We discuss the implications of our discoveries in Section \ref{discussion} and summarize our work in Section \ref{summary}.

\begin{deluxetable*}{ccccc}
\tablecaption{Data Sets Used in This Work. \label{observation_log_all}} 
\tablehead{\colhead {Instrument} & ObsID & \colhead{MJD Range} & \colhead{Energy (Wavelength)} & \colhead{Time Resolution}}
\startdata
\multicolumn{5}{c}{X-Ray All-Sky Monitoring Programs}\\
\hline
\rxte\ ASM & \nodata & 50123--55200 & 1.5--12\,keV & 90\,minutes \\
\swift\ BAT & \nodata & 53416--58565 &  15--50\,keV & 96\,minutes \\
MAXI GSC & \nodata & 55058--58565 & 2--20\,keV & 92\,minutes \\
MAXI GSC (events) & \nodata & 55058--58565 & 2--20\,keV & 50\,$\mu$s\\
\hline
\multicolumn{5}{c}{Pointed X-Ray Observations}\\
\hline
\chandra\ HRC & \dataset[14054]{http://cda.harvard.edu/chaser/viewerContents.do?obsid=14054} & 55969.37--55970.26 & 0.08--10\,keV & 16$\mu$s \\
\xmm\ PN & \dataset[0784570201]{http://nxsa.esac.esa.int/nxsa-web/\#obsid=0784570201} & 57639.93--57640.14 & 0.3--10\,keV & 30$\mu$s\\
\xmm\ PN & \dataset[0784570301]{http://nxsa.esac.esa.int/nxsa-web/\#obsid=0784570301} & 57650.32--57650.54 & 0.3--10\,keV & 30$\mu$s\\
\xmm\ PN & \dataset[0784570501]{http://nxsa.esac.esa.int/nxsa-web/\#obsid=0784570501} & 57685.91--57686.13 & 0.3--10\,keV & 30$\mu$s\\
\swift\ XRT & \dataset[00081950001]{http://www.swift.ac.uk/archive/browsedata.php?oid=00081950001&source=obs} & 57650.15--57650.55 & 0.3--10\,keV & 1.8\,ms \\
\swift\ XRT & \dataset[00081950002]{http://www.swift.ac.uk/archive/browsedata.php?oid=00081950002&source=obs} & 57685.21--57685.94 & 0.3--10\,keV & 1.8\,ms \\
\hline
\multicolumn{5}{c}{Optical Survey}\\
\hline
ASAS-SN (V-band) & \nodata & 56792--58385 & 551(83)\,nm$^a$ & 90\,s \\
ASAS-SN (g'-band) & \nodata & 58002--58483 & 480(141)\,nm$^a$ & 90\,s 
\enddata
\tablenotetext{a}{This value denotes the central wavelength of the filter. The FWHM of the filter is shown in the parentheses.}
\end{deluxetable*}

\section{Observations and Data Reduction}\label{observation}
\subsection{X-ray Monitoring Data}
We investigate the superorbital modulation using the data taken with the ASM on board \rxte, the Burst Alert Telescope (BAT) on board \swift, and the Gas Slit Camera (GSC) on board MAXI. We use the dwell data of ASM and one-satellite-orbit binned light curve of \swift\ BAT and MAXI to investigate the orbital modulation. The exposure time of the ASM dwell light curve, the time resolution of the one-satellite-orbit binned BAT and MAXI light curves, and the time coverage and the energy range of these three satellites are summarized in Table \ref{observation_log_all}. We rebin the light curves with a resolution of 1 day to trace the superorbital modulation of SMC X-1. We further utilize the photon events collected with the MAXI GSC to track the spin period evolution of SMC X-1.

The ASM consists of three proportional counterarrays with a collecting area of 90\,cm$^2$ \citep{LevineBC1996}. The energy range of the ASM is 1.5--12\,keV and can be divided into three bands, 1.5--3, 3--5, and 5--12\,keV to provide X-ray hardness information. It swept the entire sky every $\sim$90 minutes from 1996 to 2012. We exclude the data collected after MJD 55400 (2010 July) because the gain of the ASM changes significantly and the superorbital modulation is difficult to recognize \citep{LevineBC2011}. To eliminate possible contamination from the bad data points or extremely short exposures, we filter these data points with uncertainties that are 3$\sigma$ higher than the mean uncertainty. We also remove data points with background count rates that are 3$\sigma$ higher than the mean level.

The BAT onboard \swift\ is designed to trigger alerts of gamma-ray bursts. It has a large collecting area (5200\,cm$^2$) and has monitored known X-ray sources in the hard X-ray (15--150\,keV) since 2004 \citep{BarthelmyBC2005}. The entire sky can be scanned once every $\sim$96 minutes. We use the 15--50\,keV light curve provided by the hard X-ray transient monitor program \citep{KrimmHC2013}. We apply the same selection criterion on the uncertainties as for the ASM data to filter the light curve. 

MAXI is a payload mounted on the Japanese Experimental Module of the \emph{International Space Station} (\emph{ISS}). Similar to the ASM, it can monitor the entire sky in the X-ray band \citep{MatsuokaKU2009}. MAXI is equipped with two cameras: the solid-state slit camera (SSC) with a collecting area of 200\,cm$^2$ in the energy range of 0.5--12 keV, and the GSC, with a collecting area of 5350\,cm$^2$ in the energy range of 2--30 keV. We investigate the orbital and superorbital modulations of SMC X-1 with the GSC light curve provided by RIKEN, JAXA, and the MAXI team. The energy range of the archival binned light curves is 2--20\,keV. The hardness information can also be obtained because the light curves are further divided into 2--4, 4--10, and 10--20\,keV bands. We apply the same selection criterion on the uncertainties as for the ASM data to filter the light curve.

Except for the archival light curves, we search for pulsation signal and track the spin period evolution of SMC X-1 using photon events collected with MAXI.  We only use photons collected with the GSC owing to its excellent time resolution of 50\,$\mu$s \citep[20\,kHz clock counter;][]{MiharaNS2011}, which is suitable to track the evolution of accreting pulsars \citep[see, e.g.,][]{TakagiMS2016, YatabeMM2018}. The data transmission from the \emph{ISS} to the ground has three reduction modes: 64, 32, and 16\,bit, due to the limited telemetry bandwidth. We use 64\,bit data because they contain precise timing information \citep{MiharaNS2011}. The GSC consists of 12 proportional counters (GSC ID 0, 1, ..., 9, A, B) encapsulated in two mission data processors \citep{MiharaNS2011}; GSC 3, 6, and 9 are out of order, and GSC A, B stopped working before 2010 \citep[see Section 2.3 in][]{SugizakiMS2011}. Therefore, we only use data with GSC IDs of 0, 1, 2, 4, 5, 7, and 8. We extract X-ray photons in 2--20\,keV before MJD 58565 using \texttt{mxextract} from the MAXI database and correct the arrival time of each photon to the barycenter of the solar system. We extract the source photons from a 1$^{\circ}$ radius circle centered on SMC X-1. Nearly 90\% of the source photons are encircled \citep{MiharaNS2011}. 

\subsection{Pointed X-Ray Observations}
The pointed observations made with \chandra, \xmm, and \swift\ XRT are ideal benchmarks for tracking the spin period evolution and searching for the spin period with MAXI GSC events (see Section \ref{spin_period_section}). The basic information of pointed observations used in this work is summarized in Table \ref{observation_log_all}. 

Between 2001 and 2016, SMC X-1 was observed with \chandra\ 19 times. The spin period measurements in 2001--2002 are reported in \citet{Neilsen2004}. The observation in 2013 (ObsID \dataset[14054]{http://cda.harvard.edu/chaser/viewerContents.do?obsid=14054}) was made with the High Resolution Camera (HRC) with a time resolution of 10\,ms, enough to resolve the spin signal of SMC X-1. Ten other observations during 2012 and 2016 were made with the high-energy transmission grating in timed-exposure mode. We do not make use of these data due to their insufficient time resolution. We reprocess the data (ObsID \dataset[14054]{http://cda.harvard.edu/chaser/viewerContents.do?obsid=14054}) using the pipeline \texttt{chandra\_repro} in the \chandra\ Interactive Analysis of Observations (CIAO) version 4.9 with the calibration database (CALDB) version 4.7.3 \citep{FruscioneMA2006}. We correct the arrival time of each photon to the barycenter of the solar system using the \texttt{axbary} tool. We extract the source events from a circular aperture of 2\arcsec\ radius.  

Seven observations are carried out by \xmm\ where SMC X-1 is in the field of view. The spin period obtained from two 2001 observations (ObsIDs \dataset[0011450101]{http://nxsa.esac.esa.int/nxsa-web/\#obsid=0011450101} and \dataset[0011450201]{http://nxsa.esac.esa.int/nxsa-web/\#obsid=0011450201}) are reported in \citet{Neilsen2004}. The main target of the observation made in 2006 (ObsID \dataset[0311590601]{http://nxsa.esac.esa.int/nxsa-web/\#obsid=0311590601}) is Nova SMC 2005, and SMC X-1 is in the field of view. However, this observation was made during the eclipse; hence, we could not search for pulsation. In 2016, \xmm\ carried four observations (ObsIDs \dataset[0784570201]{http://nxsa.esac.esa.int/nxsa-web/\#obsid=0784570201},\dataset[0784570301]{http://nxsa.esac.esa.int/nxsa-web/\#obsid=0784570301}, \dataset[0784570401]{http://nxsa.esac.esa.int/nxsa-web/\#obsid=0784570401}, and \dataset[0784570501]{http://nxsa.esac.esa.int/nxsa-web/\#obsid=0784570501}). The PN detector is operated in timing mode, which is ideal for our purpose owing to its excellent time resolution of 30\,$\mu$s. We perform basic data reduction with the task \texttt{epproc} of the \xmm\ Science Analysis System (SAS). We filter out the time intervals with strong background flaring and correct the photon arrival time to the barycenter using the \texttt{barycen} task. The source events are extracted from 19 pixels in RAWX around the source. This corresponds to a 1\arcmin.3 $\times$ 13\arcmin.5 box on the sky. We exclude the dataset with ObsID \dataset[0784570401]{http://nxsa.esac.esa.int/nxsa-web/\#obsid=0784570401} in the following analysis, because it is made during the superorbital low state and the spin signal cannot be detected.

Twice in 2016, SMC X-1 was observed with \swift\ where the XRT was operated in the windowed timing mode that provides a high time resolution of 1.8\,ms. We utilize the task \texttt{xrtpipeline} to create cleaned level 2 event files. The image is compressed into one dimension. We choose the source events from a 60\arcsec\ wide box from SMC X-1 to ensure that $\sim$90\% of photons are collected. The vertical size corresponds to 8\arcmin\ on the sky. The barycentric correction is achieved using the tool \texttt{barycorr}. 

\subsection{Optical Light Curve}
We retrieve the optical light curve of SMC X-1 from the ASAS-SN sky patrol\footnote{\url{https://asas-sn.osu.edu/}} to independently confirm the orbital and superorbital modulations obtained by the Optical Gravitational Lensing Experiment (OGLE) in the $I$ band \citep{CoeAO2013}. The ASAS-SN has scanned the entire sky every 2--3 days in the $V$ band since 2013 and the $g$ band since 2017. SMC X-1 is monitored since May 2014. Roughly $\sim$1300 exposures in $V$-band and $\sim$900 exposures in $g$-band are used in this research. The exposure time of each snapshot is 90\,s. The limiting magnitude of each exposure varies between observations but is concentrated in $V=15.3-16.3$ and $g=15.5-16.5$, which is two magnitude fainter than the typical apparent magnitude of SMC X-1 ($V\sim 13.15$ and $g\sim13.05$). We remove those data points without magnitude measurement and filter out those data points with uncertainties that are 3$\sigma$ higher than the mean uncertainty. The observational periods of the $V$ and $g$ bands have a 1 yr overlap between MJD 58002 and 58385. The MJD range of the ASAS-SN data used in this analysis is described in Table \ref{observation_log_all}. 

\section{Analysis and Result}\label{result}
In this section, we describe the analysis technique and the result of the superorbital, orbital, and spin period evolution. The result of the superorbital modulation is described in Section \ref{superorbital_section}. We first utilize two time-frequency analysis methods to track the evolution of the superorbital modulation frequency and find a new excursion event (Section \ref{excursion_section}). Then we study the superorbital profile in detail, including the hardness ratio (HR) variability, the connection between the modulation amplitude and cycle length, and the high-/low-state flux variability (Section \ref{superorbital_profile_section} and \ref{corr_amp_cycle_length_section}). In Section \ref{spin_period_section}, we track the spin frequency evolution of SMC X-1 with both pointed X-ray observations and the MAXI GSC event files. We deeply investigate the connection between the spin frequency residual and the superorbital frequency. By tracking the phase shift of the mean longitude of the orbit, we refine the orbital ephemeris (Section \ref{orbital_ephemeris_section}). In Section \ref{orbital_profile_section}, we investigate the detailed orbital profile and the HR in different superorbital states with the MAXI GSC. Finally,  we study the orbital and superorbital modulations in the optical band in Section \ref{optical_section}.  

\begin{figure}
\includegraphics[width=0.47\textwidth]{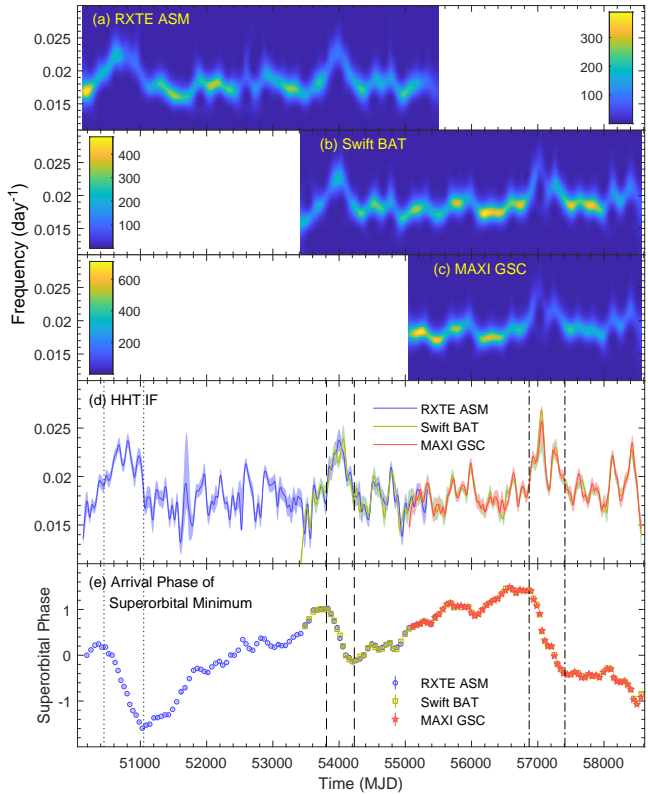} 
\caption{The WWZ spectra of SMC X-1 obtained from (a) \rxte\ ASM, (b) \swift\ BAT, and (c) MAXI GSC light curves. The color bar denotes the $Z$ value of WWZ spectra. The instantaneous frequencies obtained from the \rxte\ ASM (blue), \swift\ BAT (green), and MAXI GSC (red) light curves with the HHT are plotted in panel (d). The frequencies are smoothed, and the corresponding shaded areas denote 95\% confidence intervals. The arrival phase of the minimum of the superorbital modulation is shown in panel (e).  Three excursion epochs (see Table \ref{epoch_periods} for definition) are labeled as dotted, dashed, and dashed-dotted vertical lines.  \label{superorbital_all}}
\end{figure}

\subsection{Superorbital Modulation} \label{superorbital_section}

The variability of the superorbital period before $\sim$MJD 55500 has been carried out with the \rxte\ ASM data \citep{Hu2011, DageCC2018}.  We analyze the data collected with the \swift\ BAT and MAXI GSC to extend the time baseline to 2019 March ($\sim$MJD 58560) and track the superorbital evolution of SMC X-1 beyond the \rxte\ era. We use the weighted wavelet $Z$-transform \citep[WWZ; ][]{Foster1996} and the Hilbert-Huang transform \citep[HHT; ][]{Huang1998} to analyze the data. 

\subsubsection{Time-frequency Analysis Algorithms}\label{time_frequency_section}
The WWZ is based on the Morlet wavelet algorithm and generalized to unevenly sampled time series. The wavelet function is a sinusoidal wave convolved by a Gaussian envelope \citep{GrossmannM1984}. The wavelet of a trial frequency $\omega$ at a time $\tau$ is
\begin{equation}
F(t)=e^{i\omega(t-\tau)-c\omega^2(t-\tau)^2},
\end{equation}
where $c$ is a constant to scale the width of the Gaussian envelope. 


The HHT contains two steps: (i) decompose the light curve into intrinsic mode functions (IMFs) with empirical mode decomposition (EMD) and (ii) perform the Hilbert transform or other algorithms, e.g., direct quadrature or generalized zero-crossing, to yield the instantaneous frequency. The instantaneous frequency can be calculated from any time series, but it is physically meaningful only if the time series is an IMF. Two criteria should be satisfied for an IMF: (i) the number of extrema and the number of zero crossings must be identical or, at most, differ by one; and (ii) the local mean value defined by the average of the upper and lower envelopes must be zero \citep{Huang1998}. They are nearly impossible for the real data, and hence the EMD is proposed \citep{Huang1998}. A time series $x(t)$ can be expressed as
\begin{equation}
    x(t)=\sum_{j=1}^n c_j(t)= \sum_{j=1}^n a_j(t)e^{i\int\omega_j(t)dt},
\end{equation}
where $c_j(t)$ is the decomposed IMFs, $a_j(t)$ is the instantaneous amplitude, $\omega_j(t)$ is the instantaneous frequency of the $j$th IMF, and $n$ is the number of IMFs. In general, $n\leq\log_2 N$ where $N$ is the number of total data points in $x(t)$. Compared to the Fourier analysis, the HHT decomposes the time series into fewer components, but the amplitude and frequency of each component are functions of time.

However, a signal with a consistent modulation timescale could be decomposed into different IMFs using EMD. This ``mode-mixing'' problem is solved by the assistance of white noise \citep[EEMD;][]{Wu2009}. For each trial, a white-noise series with a finite amplitude is added to the data. We then obtain the mean of each IMF by 
\begin{equation}
    c_j=\frac{1}{M}\sum_{k=1}^M c_{jk}
\end{equation}
where $M$ is the number of total trials. A remaining problem is that the summation of IMFs is no longer an IMF. We apply a post-processing EMD to deal with this problem \citep[see][for a detailed procedure]{Wu2009}. This algorithm has been successfully used to analyze the \rxte\ ASM light curve of SMC X-1 \citep{Hu2011}.

Recently, the EEMD has been further improved by adding a pair of positive and negative white noises to optimize the completeness of the IMFs \citep{YehSH2010}. This is included in the Matlab fast EEMD package developed by the Research Center for Adaptive Data Analysis at National Central University \citep{WangYY2014}\footnote{\url{http://in.ncu.edu.tw/ncu34951/research1.htm}}. This computationally efficient algorithm allows us to perform a Monte Carlo simulation and estimate the confidence interval of the instantaneous frequency obtained with the normalized Hilbert transform. We generate $10^4$ simulated light curves based on the observed count rate plus a Gaussian-distributed noise with the standard deviation equal to the uncertainty. For each light curve, we obtain the instantaneous superorbital frequency. We then take the average of the instantaneous frequency at day.

\subsubsection{Superorbital Period Excursions}\label{excursion_section}

The resulting WWZ spectra of the \rxte\ ASM, \swift\ BAT, and MAXI GSC light curves are plotted in Figure \ref{superorbital_all} (a)--(c), while the mean instantaneous frequencies obtained with the HHT are plotted in Figure \ref{superorbital_all}(d). The mean instantaneous frequencies are smoothed by a local regression smoothing weighted by a tricube function to emphasize the structures longer than one superorbital cycle \citep{Cleveland1979, ClevelandD1988}. The 95\% confidence intervals obtained from all three observatories are plotted in Figure \ref{superorbital_all}(d). 

From the WWZ and the HHT results, we observe that the superorbital frequency evolved to $f_{\rm{sup}}\approx0.025$ ($P_{\rm{sup}}\approx40$\,days) around $\sim$MJD 57100, similar to the previous two excursion events on $\sim$MJD 50800 and $\sim$ MJD 54000. The latter two events on $\sim$MJD 54000 and $\sim$MJD 57100 are independently observed by at least two satellites. The time interval between two successful excursion events is $\sim$3150 days, implying that the excursion event is recurrent and possibly periodic. 

\begin{deluxetable}{ccc}
\tablecaption{Averaged Superorbital Modulation Period of SMC X-1 in Five Epochs. \label{epoch_periods}} 
\tablehead{\colhead{MJD Range} & \colhead{Epochs} & \colhead{Mean Period (days)}}
\startdata
50450--51050 & Excursion & 45.8(2) \\
51050--53810 & Regular & 56.9(1) \\
53810--54230 & Excursion & 45.8(3)\\
54230--56870 & Regular & 56.1(1) \\
56870--57410 & Excursion & 45.2(3)\\
\enddata
\tablecomments{The uncertainties are $1\sigma$ statistical uncertainties, which may be underestimated because the period has short-timescale variability.}
\end{deluxetable}

The EEMD is equivalent to a local dyadic filter and can be used to filter high-frequency noise. For each dataset, we combine the IMFs from the $c_5$ \citep[dominated by the superorbital modulation; see][]{Hu2011} to the residual to obtain a high-pass-filtered light curve, similar to the operation to the quasiperiodic oscillation in RE J1034+396 \citep{Hu2014}. We then get the arrival time of each superorbital minimum. The uncertainty of the arrival time of each superorbital minimum is estimated by the Monte Carlo simulation. We plot the evolution of the arrival phase with a 54.3\,day folding period in Figure \ref{superorbital_all}(e). The arrival times of superorbital minima determined with the \rxte\ ASM, \swift\ BAT, and MAXI GSC are listed in Table \ref{all_superorbital_minimum} of Appendix \ref{appendix_superorbital}. We identify three major excursion epochs and two regular epochs from the phase evolution. We define the excursion epoch as the period when the arrival phase of the superorbital minimum decreases more than one cycle. Two regular epochs are then defined as the intervals between these three excursion epochs. The mean superorbital periods in these five epochs are listed in Table \ref{epoch_periods}. The mean superorbital periods during the major excursion epochs are as short as $\sim45$\,days. In contrast, the mean superorbital period is $\sim56$\,days in the regular epochs. The phase evolution of these two regular epochs is not stable, but the patterns are similar. A humpy pattern can be seen during the middle $\sim$1000 days of the regular epoch, and a few minor excursions that contain to to five superorbital cycles with $P_{\rm{sup}}\lesssim50$\,days occasionally occur.  The phase of the last few cycles after $\sim$MJD 58000 drops significantly. This coincides with the variation toward shorter superorbital periods (see Figure \ref{superorbital_all}(a)--(d)). In the current stage, we could not conclude whether it indicates a new major period excursion event or a minor excursion event similar to those near $\sim$MJD 52000, and $\sim$MJD 56000.  

\begin{figure}
\includegraphics[width=0.47\textwidth]{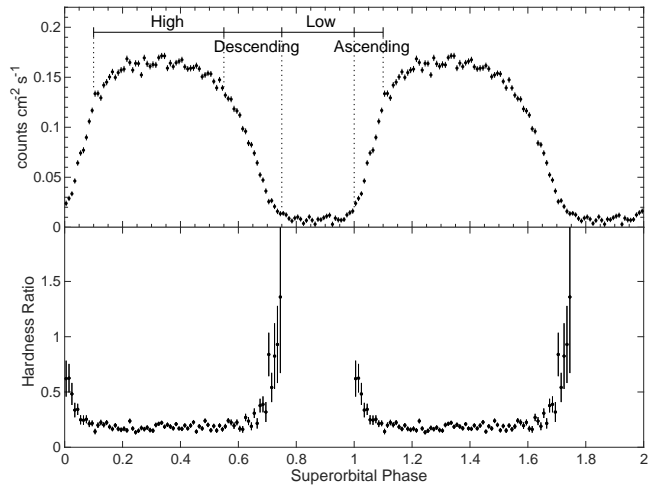} 
\caption{(Top) MAXI GSC 2--20\,keV light curve folded with the superorbital phase defined with the HHT. (Bottom) HR vs.~superorbital phase. The bin size is $1/100$ cycle, where HR values in superorbital phase $0.75\leq\phi_{\rm{sup}}<1.0$ are not presented. \label{hardness_superorbital}}
\end{figure}

\subsubsection{Superorbital Profile} \label{superorbital_profile_section}
The superorbital profile can be obtained according to the instantaneous phase derived in the HHT. The profile shapes obtained from the \rxte\ ASM, \swift\ BAT, and MAXI GSC are fully consistent with each other, implying that the mechanism is nearly energy-independent. The superorbital profile observed with the MAXI GSC is shown in Figure \ref{hardness_superorbital}. We assign the start of the ascending as phase zero. The profile shows an ascending lasting $\sim$0.1 cycles (from phase 0--0.1), a high state at phase 0.1--0.55, a descending at phase 0.55--0.75, and a low state at phase 0.75--1.0. 

The spectral variability can be obtained with the variability of the HR. Utilizing the \rxte\ ASM data, it has been suggested that the emission during the low state is dominated by hard X-rays \citep{Trowbridge2007}.  In this research, we investigate the detailed HR variability with the MAXI GSC. The HR is defined as
\begin{equation}\label{hr_def}
\rm{HR}=\frac{\rm{ch2}-\rm{ch1}}{\rm{ch2}+\rm{ch1}}\rm{,}
\end{equation}
where ch1 is the count rate in 2--4\,keV, and ch2 is the count rate in 4--10\,keV. A high HR value implies that the emission is dominated by hard X-rays. We first create a superorbital profile in each band by dividing a superorbital cycle into 100 bins. Then we calculate the HR value in each bin using Equation \ref{hr_def}. The X-ray photons collected during the eclipse are excluded. The superorbital and the HR profiles are shown in Figure \ref{hardness_superorbital}. We find that the HR values fluctuate significantly and have extremely large uncertainties in the superorbital phase $0.75\leq\phi_{\rm{sup}}<1.0$. We remove the corresponding data points because SMC X-1 is unlikely to be detectable during the superorbital low state. The HR is stable at $\sim0.25$ during the high state, while it gradually decreases in the ascending state and gradually increases in the descending state. This is the first time we witness the change of HR in the transition with X-ray monitoring data. This suggests a possible variability of X-ray absorption caused by a gradual change of the optical depth \citep{Inam2010}.

\begin{figure}
\includegraphics[width=0.47\textwidth]{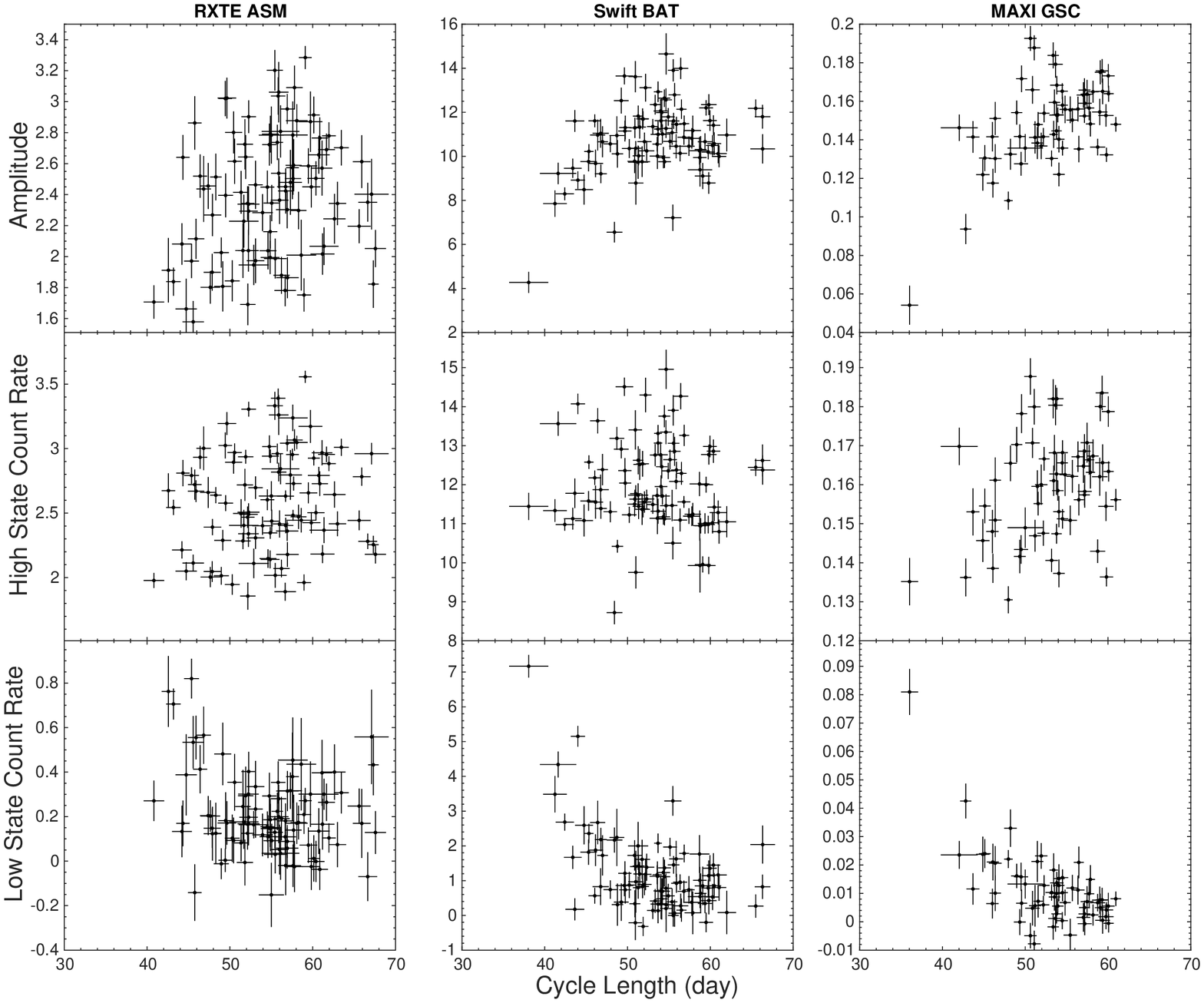} 
\includegraphics[width=0.47\textwidth]{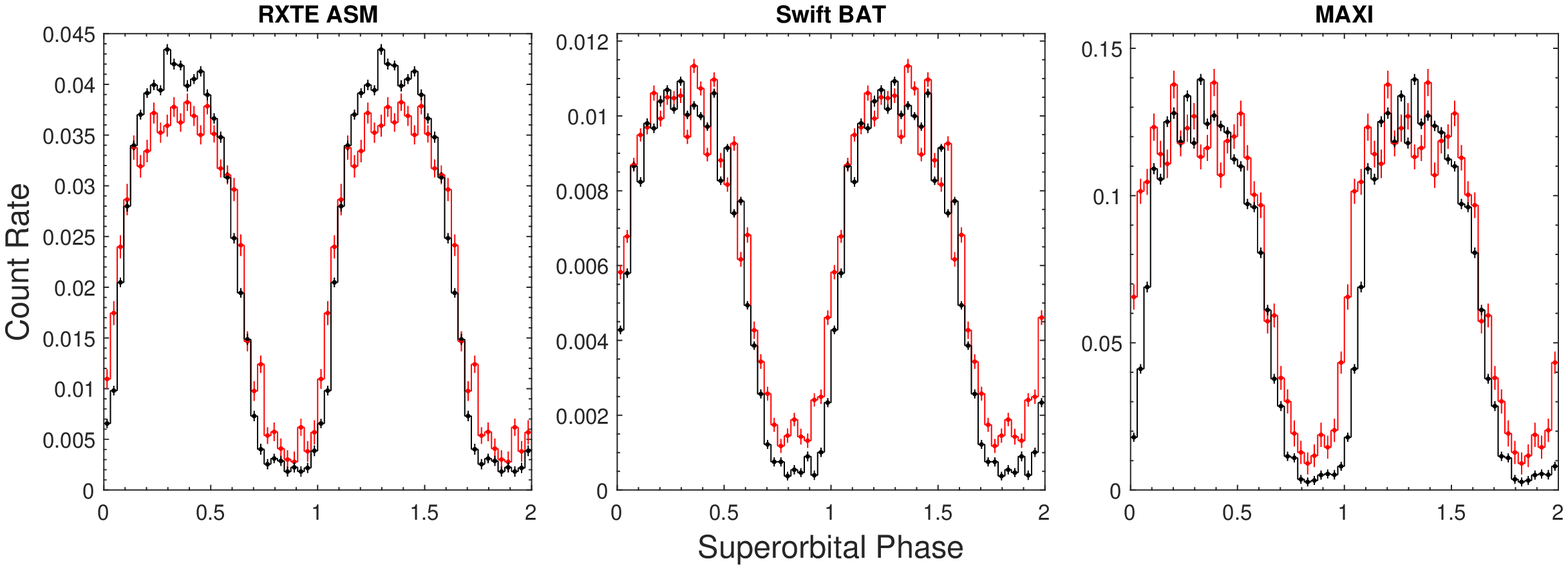} 
\caption{(Top) Modulation amplitude, the high-state count rate, and the low-state count rate versus the superorbital cycle length. (Bottom) Superorbital profiles in excursion epochs (red) and regular epochs (black). \label{fractional_variability}}
\end{figure}

\subsubsection{Correlation between Superorbital Amplitude and Cycle Length} \label{corr_amp_cycle_length_section}

It is suggested that the superorbital modulation amplitude may marginally correlate with the superorbital cycle length \citep{Hu2011}. We reexamine this correlation with new \swift\ BAT and MAXI GSC datasets. The cycle length is obtained from the HHT phase, where the uncertainty can be estimated by the Monte Carlo simulation in Section \ref{excursion_section}. We also try to define the cycle length from the time interval between two neighboring superorbital minima or maxima, and the result is fully consistent. In each cycle, we obtain an amplitude from the difference between the averaged high-state count rate and the averaged low-state count rate. This is a more detailed analysis than the rms amplitude presented in \citet{Hu2011} because the variability of high- and low-state fluxes could be originated from different mechanisms. 

The top panel of Figure \ref{fractional_variability} shows the superorbital amplitude and high-/low-state count rates versus superorbital cycle length. We find that the correlation between modulation amplitude and the cycle length is not well determined in \rxte\ ASM data but can be seen in the \swift\ BAT and MAXI GSC data. The Pearson's linear correlation coefficient of \swift\ BAT result is $r=0.34$, with a null hypothesis probability of $p=9.1\times10^{-4}$.  The correlation obtained from the MAXI GSC is $r=0.57$, with a null hypothesis probability of $p=8.4\times10^{-7}$. We examine the high- and low-state count rates and find that the correlation is dominated by the variability of the count rate in the low state. The correlation coefficient between the cycle length and the low-state count rate in the BAT data set is $r=-0.52$ with $p=6.5\times10^{-8}$, while it is $r=-0.64$ with $p=1.5\times10^{-6}$ for the MAXI GSC. The Pearson's correlation assumes a linear relationship that may not be suitable for our dataset. Therefore, we use the nonparametric Kendall's correlation coefficient ($\tau_{\rm{k}}$) to further test the correlation between the cycle length and the low-state count rate. For the BAT data set, we obtain $\tau_k=-0.27$ with $p=1.7\times10^{-4}$, while for the MAXI data, we obtain $\tau_k=0.35$ with $p=7.4\times10^{-5}$. Considering that the correlation does not take the uncertainties into account, we perform $10^6$ times Monte Carlo simulations. In each trial, we generate a fake dataset of cycle length and amplitude. For each data point, we assign the values by generating two Gaussian-distributed random numbers centered on the observed values of cycle length and amplitude and $\sigma$ equal to the uncertainties. The resulting distribution of the correlation coefficient is well concentrated near the observed value with a limited deviation. The mean value of $\tau_k$ between the cycle length and the low-state count rate in the \swift\ BAT dataset is $-0.27$ with a standard deviation of $0.03$, while the mean and standard deviation values for the MAXI data set are $-0.34$ and $0.04$, respectively.

However, the correlation could be dominated by a few outliers. We use the bootstrap resampling technique with $10^7$ simulations to test whether the observed correlation is biased \citep{Efron1979}. The mean value of $\tau_k$ between the cycle length and the low-state count rate of the \swift\ BAT data is $-0.28\pm0.07$, where the probability that showing a positive correlation is $3.5\times10^{-4}$. The mean value of $\tau_k$ of the MAXI GSC data is $-0.36\pm0.08$ with a probability of $7.3\times10^{-5}$ that showing $\tau_k>0$. This result indicates that the negative correlation is not accidental.  In contrast, the ASM data set has a mean correlation coefficient of $-0.11$ with a standard deviation of $0.07$, and roughly 8\% of the resampled sets result in a positive correlation, indicating a marginal anticorrelation.

\begin{figure}
\includegraphics[width=0.47\textwidth]{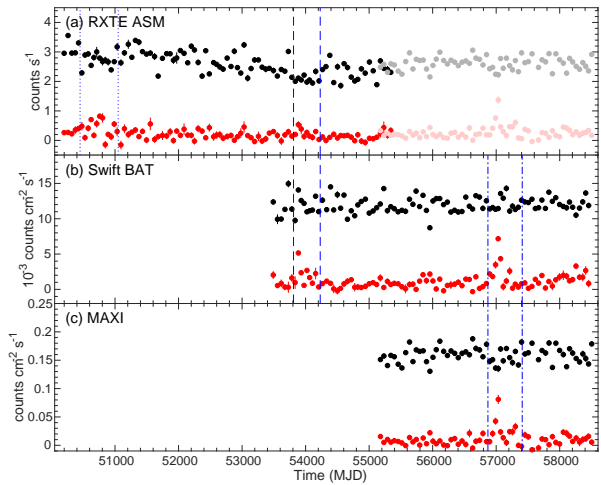} 
\caption{Mean count rates for the superorbital high state (black) and low state (red) observed with the \rxte\ ASM, \swift\ BAT, and MAXI GSC. The MAXI count rate is converted to the \rxte\ ASM count rate and stacked in panel (a) with corresponding light colors. The first, second, and third superorbital excursion epochs are denoted by dotted, dashed, and dashed-dotted vertical blue lines, respectively. \label{high_low_flux}}
\end{figure}

The true correlation between the cycle length and the low-state count rate may not be linear, although the linear correlation coefficient is significant. From the \swift\ BAT and MAXI observations, we note that the low-state count rate is nearly a constant for $P_{\rm{sup}}\gtrsim50$\,days.  Therefore, the correlation could be dominated by the low-state count rate with $P_{\rm{sup}}\lesssim 50$\,days, corresponding to the superorbital excursion events. Similar behavior can also be seen in the ASM data, though it is less obvious. To examine whether there is any difference in superorbital profile between regular and excursion epochs, we fold the light curve in both epochs (see bottom panel of Figure \ref{fractional_variability}). The valley in the superorbital profile during the excursion epochs is shallower and narrower than that during the regular epochs. In contrast, the high-state count rate of the superorbital modulation profile in the excursion epochs is fully consistent with that in regular epochs, except for the \rxte\ ASM. To verify if the flux has long-term variability, we plot the evolution of the high- and low-state count rate in Figure \ref{high_low_flux}. The energy ranges of these three data sets are quite different; hence, the count rate and the amplitude cannot be directly compared with each other. We estimate the \rxte\ ASM count rate after MJD 55400 by using of the MAXI GSC count rate. We fit the \rxte\ ASM count rate and MAXI GSC count rates in the overlapping epoch between MJD 55058 and MJD 55400 with a straight line and convert the MAXI count rate into the \rxte\ ASM count rate with the best-fit parameters.  

The high-state count rate seems to drop during the first superorbital excursion event. Then, the count rate decays gradually until the second excursion. This trend could cause the high-state count rate difference between the superorbital profile in regular and excursion epochs with the ASM. The low-state count rate likely increases during the first excursion epoch and then fluctuates around zero. However, we could not exclude the possibility that the drop of the high-state count rate is a coincidence caused by a sudden flux increase event around MJD 51000. The high-state count rate seems stable after MJD 54000. In comparison, the low-state count rate increases during the excursion epochs. This suggests that the observed correlations in Figure \ref{fractional_variability} are dominated by the variability of the low-state count rate. 

\begin{deluxetable}{cccl}
\tablecaption{Spin Frequency of SMC X-1 Measured with Pointed X-Ray Observations. \label{observation_log}} 
\tablehead{\colhead{MJD} & \colhead{Observatory} & \colhead{Spin Frequency (Hz)}}
\startdata
55969.81 & \chandra\  & 1.425752(7) \\
57640.04 & \xmm\ & 1.42941(3) \\
57650.35 & \swift\  & 1.4294(5) \\
57650.43 & \xmm\ & 1.42943(3) \\
57685.58 & \swift\ & 1.429510(6) \\
57686.02 & \xmm\ & 1.42951(3) \\
\enddata
\end{deluxetable}

\subsection{Spin Period Evolution}\label{spin_period_section}

In this section, we search for the spin frequency with several new observations made with \chandra, \swift\ XRT, and \xmm, as well as the photon events collected with the MAXI GSC. We first search for spin signal in pointed X-ray observations with the latest reported orbital ephemeris \citep{Falanga2015}. These measurements are used as cursors to perform a three-dimensional period search for MAXI GSC data (Section \ref{2dsearch_section}). We then use the measured 90$^{\circ}$ mean longitude ($T_{\pi/2}$, which is assumed to be the mid-eclipse time $T_{\rm{ecl}}$), to refine the orbital ephemeris. Finally, we update the measurements from both the pointed observation and the MAXI GSC.

\subsubsection{Algorithm of Spin Period Search}

We employ the $Z_2^2$ test to search for the spin period owing to the double-peaked pulse profile of SMC X-1 \citep{Buccheri1983}. The $Z_2^2$ value for a trial frequency is
\begin{equation}
Z_2^2=\left( \frac{2}{N_i} \right)\sum_{k=1}^2 \left[ \left( \sum_{j=1}^{N_i} \cos k\phi_j \right)^2 + \left( \sum_{j=1}^{N_i} \sin k\phi_j \right)^2 \right],
\end{equation}
where $N_i$ is the total number of photon events, and $\phi_j$ is the phase of each photon calculated from the trial frequency. Assuming a pulsar with a steady spin-up rate over the time window for the spin period measurement, the spin frequency $\nu(t)$ can be written as
\begin{equation}
\nu(t)=\nu_0+\dot{\nu}(t-T_0),
\end{equation}
where $T_0$ is the zero epoch of the spin period measurement, $\nu_0$ is the spin frequency at $T_0$, $\dot{\nu}$ is the time derivative of the spin frequency, and $t$ is the arrival time of a photon. In a binary system, the orbital Doppler effect should be taken into account. Therefore, the phase can be calculated as
\begin{equation}
\begin{aligned}
\phi_j=&\nu_0\cdot(t_j-T_0)+\frac{1}{2}\dot{\nu}\cdot(t_j-T_0)^2\\
&-\nu(t_j)\cdot\frac{a\sin i}{c} \left[ \cos(2\pi\phi_{orb,t_j})- \cos(2\pi\phi_{orb,T_0}) \right]\\
&-\nu(t_j)\cdot\frac{a\sin i}{2c}e\left[\sin(4\pi\phi_{orb,t_j}-\Omega)- \sin(4\pi\phi_{orb,T_0}-\Omega) \right]\\
\end{aligned}
\end{equation}
where $a$ is the semi-major axis of the binary orbit, $i$ is the inclination angle, $c$ is the speed of light, $e$ is the eccentricity, $\Omega$ is the periastron angle, and $\phi_{orb,T_0}$ and $\phi_{orb,t_j}$ are the orbital phases at $t_j$ and $T_0$. For pointed X-ray observations, the $\dot{\nu}$ term can be ignored owing to the short exposure. The best-determined spin period of all of the pointed observations with the updated orbital ephemeris (see Section \ref{2dsearch_section}) is summarized in Table \ref{observation_log}. 

\begin{figure}
\includegraphics[width=0.45\textwidth]{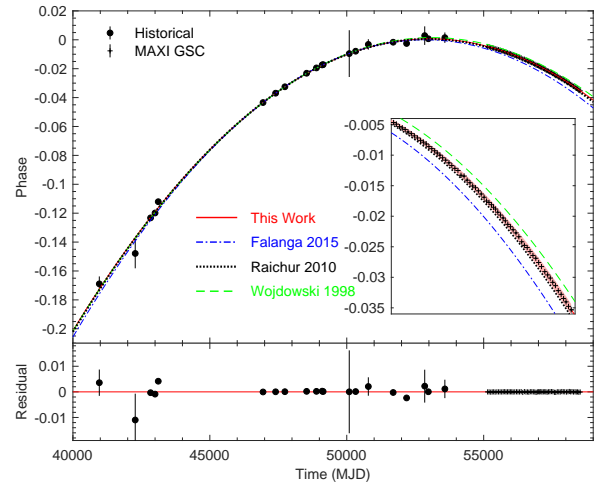} 
\caption{Arrival of the 90$^{\circ}$ mean longitude ($T_{\pi/2}$) of SMC X-1. The best-fit quadratic ephemeris and three historical ephemerides are plotted \citep{Wojdowski1998, RaichurP2010, Falanga2015}. The embedded frame is the zoom-in view of the evolution after MJD 55000. The 1-$\sigma$ uncertainty interval of the best-fit model is shown as the red shaded area. The lower panel shows the residual.   \label{toa_orbital}}
\end{figure}

\subsubsection{Refinement of the Orbital Ephemeris}\label{orbital_ephemeris_section}
The photon events collected with the MAXI GSC provide a unique opportunity to track the spin and orbital period evolution in detail. During the entire $\sim$9 yr time span, the GSC collected $\sim$1.2 million X-ray photons in 2--20\,keV within 1$^{\circ}$ of SMC X-1. We divide the observations into 65 segments, where each of them contains a full superorbital cycle.  The incomplete cycles at the boundaries are neglected. In each segment, we search for the spin signal using photons collected during noneclipse epochs and $\phi_{\rm{sup}}=0.1$--$0.6$ to achieve a good signal-to-noise ratio. The total number of photons used in this analysis is $\sim0.6$ million, while the number of photons used in each segment varies between $\sim$5000 and $\sim$18000.

We first search for the best triplet of $\nu$, $\dot{\nu}$, and $T_{\pi/2}$ in each superorbital cycle by assuming a circular orbit and freeze other orbital parameters to the values in \citet{Falanga2015}. We start our search based on the segment between MJD 55952 and 56000 because it contains a precise spin measurement with \chandra. We search for frequencies near the best-determined value and $\dot{\nu}$ from $1\times10^{-11}$\,s$^{-2}$ to $4\times10^{-11}$\,s$^{-2}$. The oversampling factor is set to 10, i.e., 10 trial $\nu$ within each peak with a Fourier width of $\delta\nu=1/2T$ where $T$ is the time span of the segment. We also use the same oversampling factor for $\dot{\nu}$, i.e., 10 trial $\dot{\nu}$ in the range of $\delta\dot{\nu}=1/T^2$. For $T_{\pi/2}$, we use a 10\,s resolution to search for the best values of $\pm 2000$\,s (corresponding to an orbital phase shift of $0.006$) around the expected value predicted by \citet{Falanga2015}. We then extend the search to all segments.

We find that the difference between the measured $T_{\pi/2}$ and the expected value increases from $\sim400$ to $\sim1600$\,s during the MAXI epoch. This implies a possible overestimate of the orbital period derivative in the orbital ephemeris. We therefore combine the historical measurements and the new $T_{\pi/2}$ values to refine the orbital ephemeris.

The orbital phase evolution can be written as
\begin{equation}
\Delta\phi_{\rm{orb}}=(\nu_{\rm{fold}}-\nu_{\rm{orb}})\cdot(t-T_{\pi/2,0})-\frac{1}{2}\dot{\nu}_{\rm{orb}}(t-T_{\pi/2,0})^2\rm{,}
\end{equation}
where $\nu_{\rm{fold}}$ is the estimated orbital frequency at $T_{\pi/2,0}$, $\nu_{\rm{orb}}$ is the corrected orbital frequency, and $\dot{\nu}_{\rm{orb}}$ is the time derivative of the orbital frequency. The orbital phases collected in \citet{Falanga2015} and the orbital phase calculated from the MAXI GSC data are presented in Figure \ref{toa_orbital}. We fit the phase evolution with a quadratic curve using the robust statistic, which is designed to minimize the effect of outliers or influential observations by iteratively reweighting the least squares with a bisquare weighting function \citep{HollandW1977, Huber1981}. The best-fit parameter is shown in Table \ref{fit_orbit_result} .A statistic of $\chi^2_{\nu}=14$ with a degree of freedom of 82 is obtained. We scale the measured 1$\sigma$ uncertainty by multiplying the value with the square root of $\chi_{\nu}^2$ as a conservative estimate (see Table \ref{fit_orbit_result}).  The best-fit model is plotted in Figure \ref{toa_orbital}. We plot the result from \citet{Wojdowski1998}, \citet{RaichurP2010}, and \citet{Falanga2015} for reference. All ephemerides seem to describe the data well, except for the MAXI observations.  Our updated model is consistent with the ephemeris proposed by \citet{RaichurP2010} at a 1$\sigma$ level but deviates from that proposed by \citet{Wojdowski1998} and \citet{Falanga2015}.

\begin{deluxetable}{ll}
\tablecaption{Orbital Parameters and Corresponding 1$\sigma$ Uncertainties of SMC X-1. \label{fit_orbit_result}} 
\tablehead{\colhead{Parameter} & \colhead{Value} }
\startdata
$T_{\pi/2,0}$ & MJD 52846.6913(2)\\
Orbital period ($P_{\rm{orb}}$) & 3.8919297(2) days\\
$\dot{P}_{\rm{orb}}/P_{\rm{orb}}$ & $-3.380(6)\times10^{-6}$\,yr$^{-1}$\\ 
\enddata
\end{deluxetable}

\begin{figure}
\begin{minipage}{0.45\linewidth}
\includegraphics[width=1.11\textwidth]{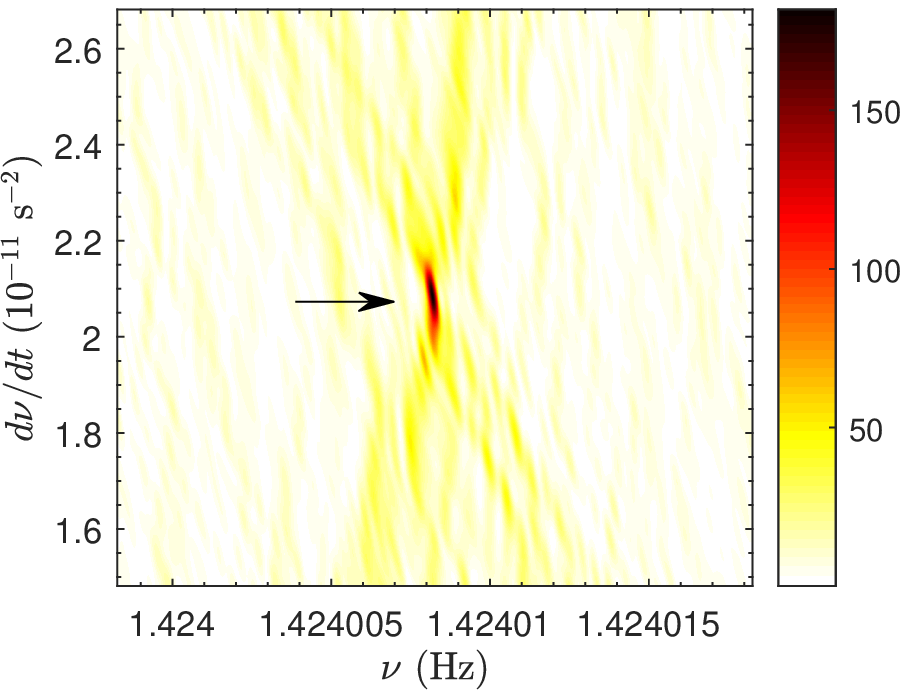}
\includegraphics[width=1.11\textwidth]{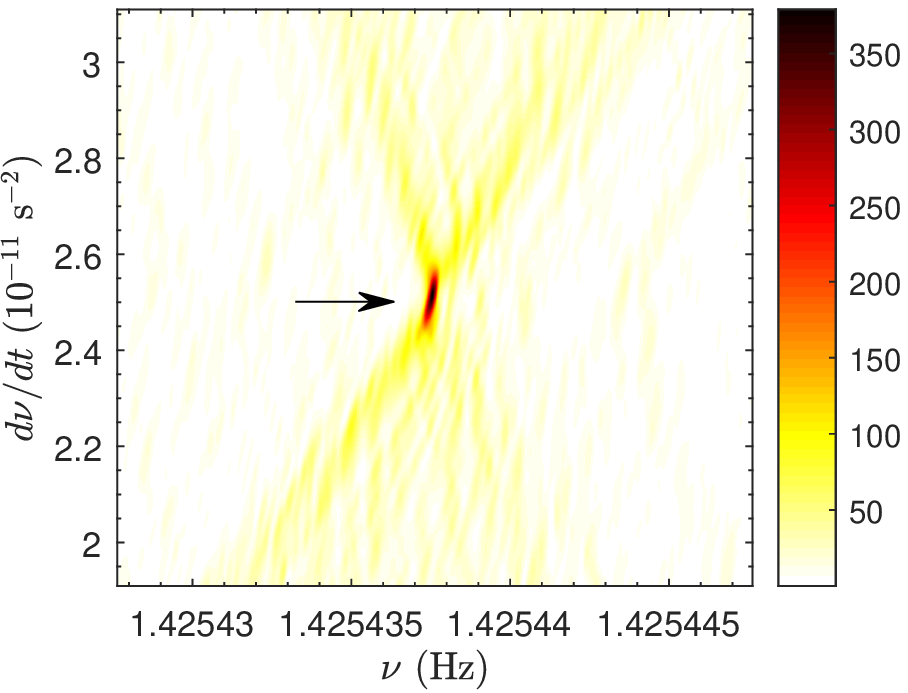}
\includegraphics[width=1.11\textwidth]{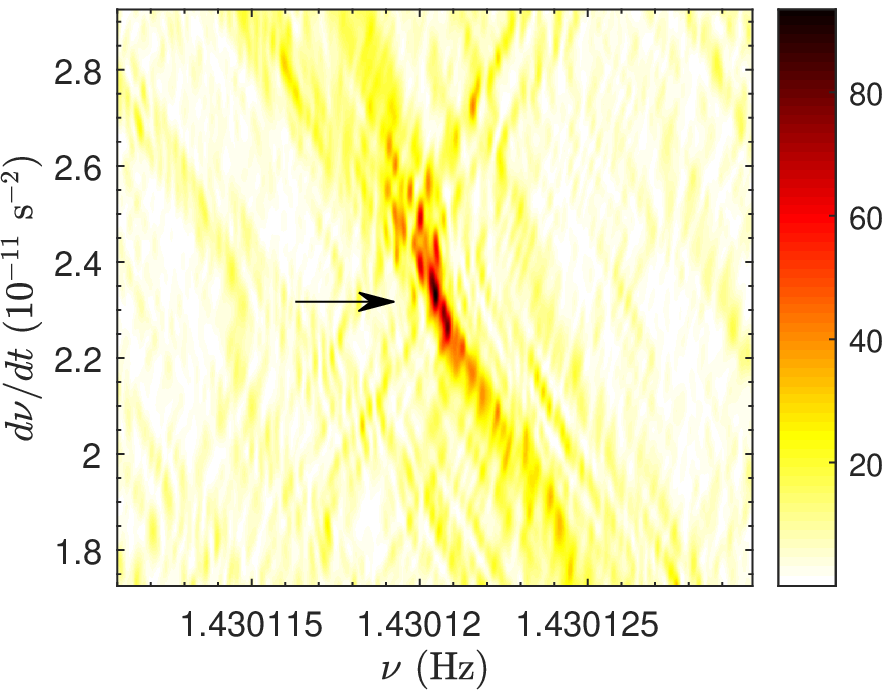}
\end{minipage}
\hspace{0.0cm}
\begin{minipage}{0.45\linewidth}
\includegraphics[width=1.0\textwidth]{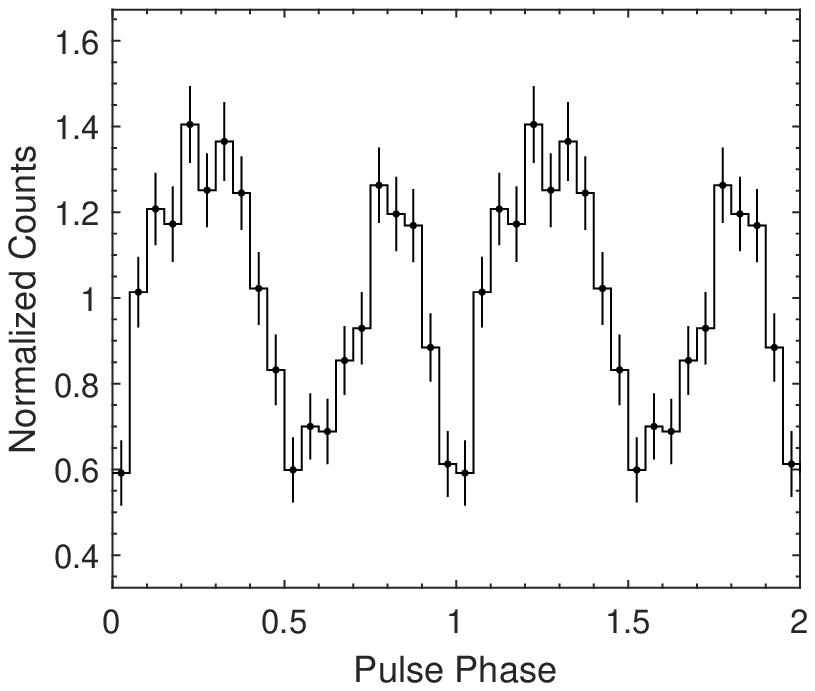}
\includegraphics[width=1.0\textwidth]{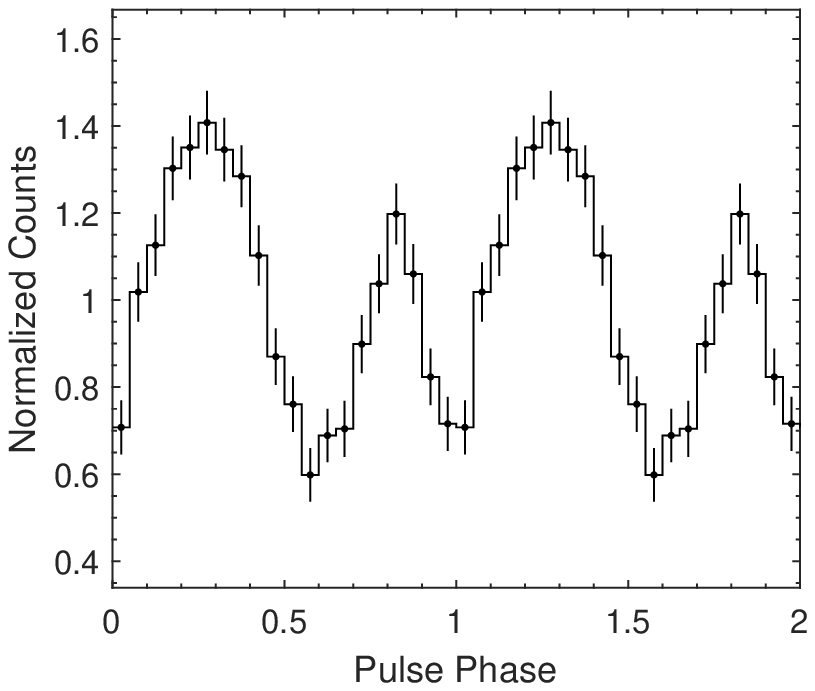}
\includegraphics[width=1.0\textwidth]{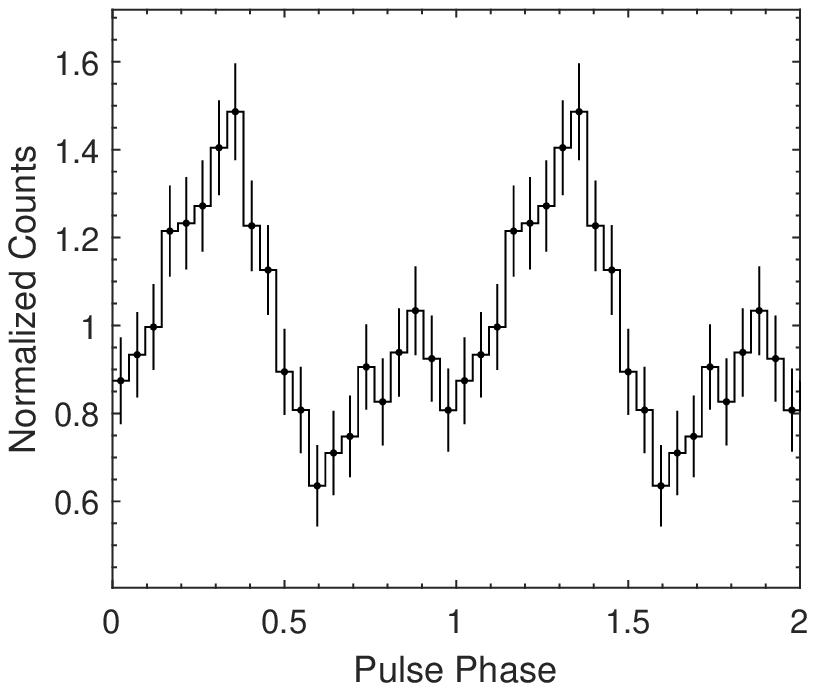}
\end{minipage}
\caption{(Left) Two-dimensional $Z_2^2$ spectra near the best-determined spin frequency of SMC X-1 with MAXI GSC data in superorbital cycles 91, 103, and 144 (top to bottom). The arrow indicates the best-determined peak on the $\nu$-$\dot{\nu}$ plane. (Right) Corresponding normalized 2--20\,keV pulse profile. Only photons collected between superorbital phases 0.1 and 0.6 are used in this analysis (see main text). \label{2dsearch_maxi}}
\end{figure}

\subsubsection{Two-dimensional Period Search with MAXI GSC}\label{2dsearch_section}

\begin{figure*}
\includegraphics[width=0.98\textwidth]{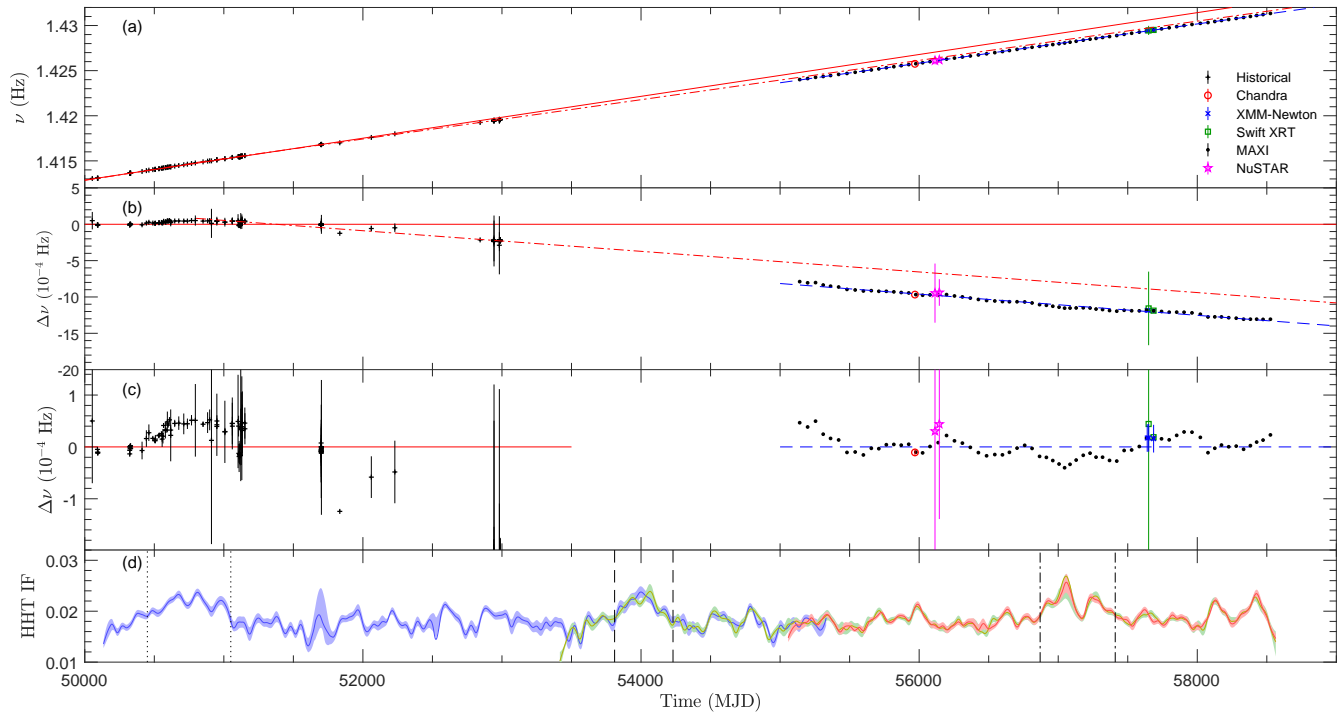} 
\caption{(a) Spin frequency evolution of SMC X-1 after MJD 50000. The red solid line denotes the best-fit linear model with data points before MJD 53000. The red dashed-dotted line denotes the best-fit linear model with data points between MJD 50800 and MJD 53000.  The blue dashed line denotes the linear model with data points collected after MJD 55000. (b) Frequency residual after subtracting the best-fit linear model with data before MJD 53000. (c) Frequency residual after separately fitting the data in MJD 50000-53000 and MJD 55000-59000 with straight lines. (d) Instantaneous superorbital frequencies obtained with the \rxte\ ASM, \swift\ BAT, and MAXI GSC. The color notations and auxiliary lines are defined in the same manner in Figure \ref{superorbital_all} \label{spin_residual}}
\end{figure*}

We then search for values of $\nu$, $\dot{\nu}$, $T_{\pi/2}$, $\Omega$, $e$, and $a\sin i/c$ to maximize the $Z_2^2$ value using the Nelder-Mead algorithm in each segment \citep{NelderM1962}. From the best-determined value of the parameters in all segments, we find that $a\sin i/c$ is well constrained as $53.48(2)$\,lt-s. A finite eccentricity $e=0.0009(3)$, and an $\Omega=10(40)^{\circ}$ are necessary. We freeze them and search for $\nu$ and $\dot{\nu}$ by using a two-dimentional $Z_2^2$ test. All of the spin period measurements obtained with MAXI are listed in Table \ref{maxi_spin_period} of Appendix \ref{appendix_spin}. The uncertainties of $\nu$ and $\dot{\nu}$ are conservatively estimated from the width of the peak. The best-determined $Z_2^2$ values vary between $\sim$36 and $\sim$380. Three examples of the two-dimensional $Z_2^2$ power spectra and the corresponding pulse profiles are shown in Figure \ref{2dsearch_maxi}. Most of the segments show a double-peaked profile in the 2--20\,keV band where the peak height and valley depths are highly variable. On the other hand, a few segments (e.g., superorbital cycle 144) show a weak secondary peak similar to panel (f) of Figure 3 in \citet{PikeHB2019}.


We plot all of the spin period measurements after MJD 50000, including the historical measurements from \citet{Inam2010} and \citet{Neilsen2004}, \nustar\ data from \citet{PikeHB2019}, and our measurements in Figure \ref{spin_residual}. The spin period of SMC X-1 increases monotonically, although short-term spin-down is possible \citep{Inam2010}. The spin-up rate is, in contrast, highly variable. Compared to the long-term trend, SMC X-1 could have a higher spin-up rate during the first superorbital excursion epoch. This could be alternatively interpreted as the accretion torque having a putative change near the midpoint of the superorbital excursion or just a coincidence \citep{DageCC2018}. By dividing the historical measurements into two segments according to MJD 50800, the spin-up rate changes from $2.816(9)\times10^{-11}$\,s$^{-2}$ to $2.520(6)\times10^{-11}$\,s$^{-2}$ (see Table \ref{fit_spin_up_rate}). We find that neither the linear model nor the two-segment model can describe the spin period measurement after MJD 55000. Figure \ref{spin_residual}(a) shows the spin period measurements, while the residual after subtracting the best-fit linear trend using data points before MJD 53000 is plotted in Figure \ref{spin_residual}(b). We further fit the measurements after MJD 55000 with a straight line, yielding a spin-up rate of $\dot{\nu}=2.515(3)\times10^{-11}$\,s$^{-2}$. This is consistent with the measurement from segment 50800--53000 but the spin frequency trends could not be connected well (see Figure \ref{spin_residual}(b)). We suggest a possible torque variability during the gap between \rxte\ and MAXI measurements, in which the second superorbital excursion event occurs. By fitting the data before MJD 53000 and after MJD 55000 separately, the corresponding residuals are plotted in Figure \ref{spin_residual}(c). The instantaneous superorbital frequency is shown in panel (d) for reference.

\begin{figure}
\includegraphics[width=0.47\textwidth]{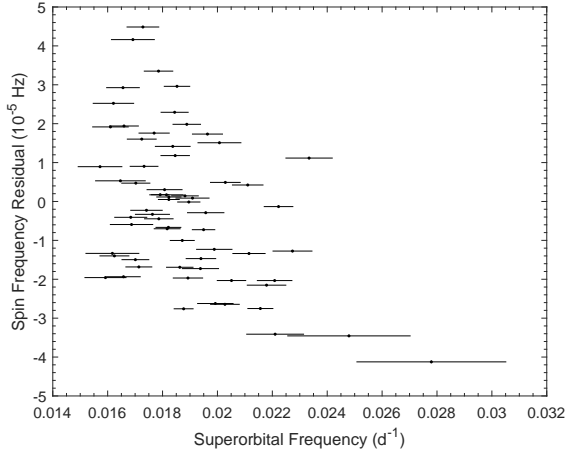} 
\caption{Spin period residual versus superorbital frequency.  \label{corr_spin_superorbital}}
\end{figure}

The spin frequency evolution obtained with the MAXI GSC shows no positive correlation with the superorbital frequency. Instead, the minimum coincides with the excursion epoch of the superorbital modulation frequency. A correlation analysis between the frequency residual and the instantaneous superorbital frequency suggests a weak anticorrelation with $\tau_k=-0.23$ and $p=8\times10^{-3}$ (Figure \ref{corr_spin_superorbital}). We use the bootstrap algorithm to perform $10^6$ simulations and obtain $\tau_k=-0.29\pm0.08$ with a probability of $1.2\times10^{-3}$ showing a positive $\tau_k$. This apparent correlation could be dominated by the of a sudden change in the spin frequency during the major superorbital excursion. By dividing the spin frequency measurement into two segments according to $\sim$MJD 57050, the spin-up rate changes from $2.489(5)\times10^{-11}$\,s$^{-2}$ to $2.544(7)\times10^{-11}$\,s$^{-2}$.



\begin{deluxetable}{lc}
\tablecaption{Local Spin-up Rate of SMC X-1. \label{fit_spin_up_rate}} 
\tablehead{\colhead{MJD Range} & \colhead{$\dot{\nu}$ (10$^{-11}$\,s$^{-2}$)} }
\startdata
50098--52988 & 2.684(3)\\
50098--50800 & 2.816(9)\\
50800--52988 & 2.520(6)\\
55141--58526 & 2.515(3)\\
55141--57050 & 2.489(5)\\
57050--58526 & 2.544(7)
\enddata
\end{deluxetable}

\begin{figure}
\includegraphics[width=0.47\textwidth]{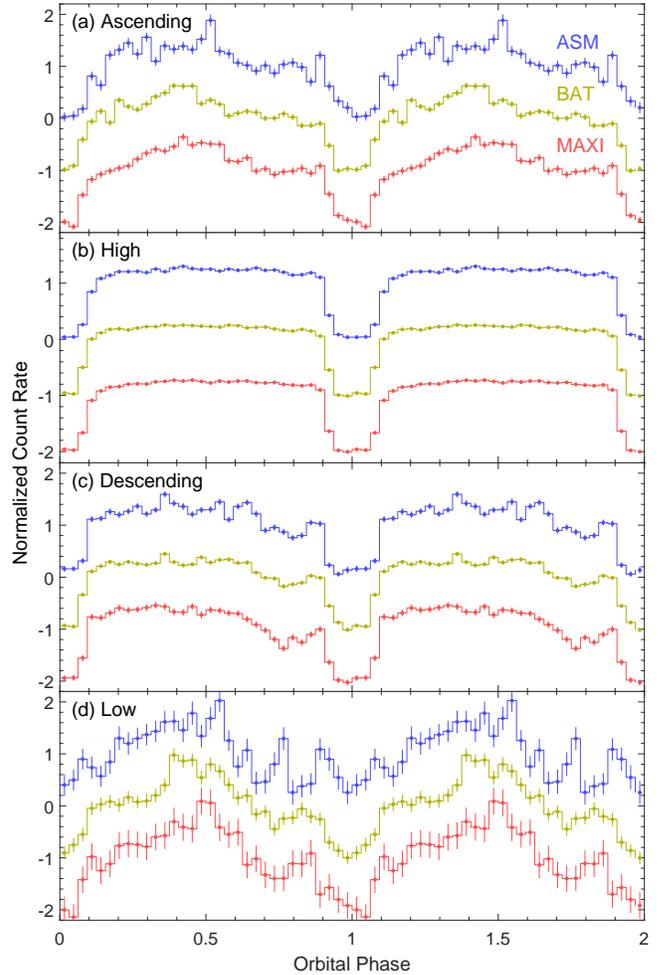} 
\caption{Orbital profile of SMC X-1 observed in (a) the ascending state, (b) the high state, (c) the descending state, and (d) the low state. Profiles obtained from the \rxte\ ASM, \swift\ BAT, and MAXI GSC are ordered from top to bottom with an offset of $-1$ in each panel. \label{orbital_all}}
\end{figure}

\subsection{Orbital Profile Variability}\label{orbital_profile_section}
The complex and variable orbital profile of SMC X-1 provides hints of the emission and absorption geometries of this system. \citet{Trowbridge2007} discovered a ``bounce-back'' feature during the eclipse with a softening and a sinusoidal profile during the noneclipse phases by observing the averaged orbital profile with the \rxte\ ASM. \citet{Hu2013} further divided the ASM data into four superorbital states and found a possible hint of pre-eclipse dips. We reexamine the orbital profile variability with respect to the superorbital phase using \swift\ BAT and MAXI GSC data. The orbital profiles in the four superorbital states are shown in Figure \ref{orbital_all}. The bounce-back feature cannot be seen. A sinusoidal-like modulation can be observed in the superorbital ascending, descending, and low states, but the X-ray flux during the post-eclipse phases ($\phi_{\rm{orb}}\sim$0.1--0.5) is likely higher than that during the pre-eclipse phases ($\phi_{\rm{orb}}\sim$0.5--0.9). Moreover, a dip-like feature at $\phi_{\rm{orb}}\sim$0.7--0.8 can be observed with all three satellites, especially in the superorbital descending state.  

\begin{figure}
\includegraphics[width=0.47\textwidth]{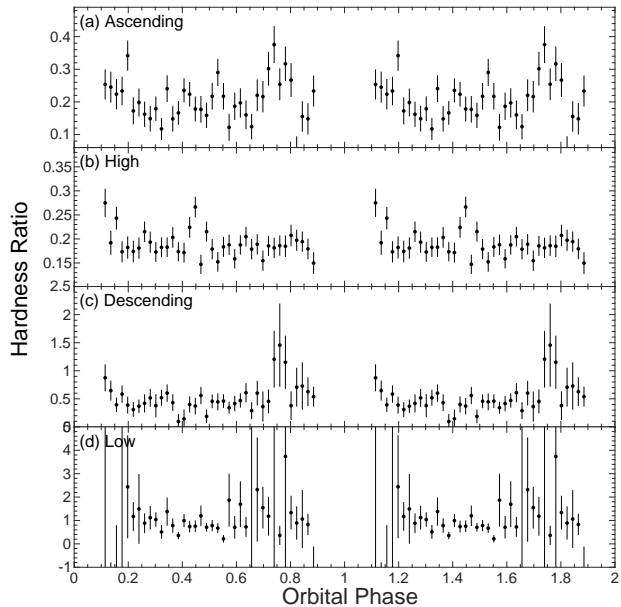} 
\caption{The HR versus orbital phase of SMC X-1 observed in (a) the ascending state, (b) the high state, (c) the descending state, and (d) the low state. \label{orbital_hardness}}
\end{figure}

We calculate the HR profile using the definition in Equation \ref{hr_def} for the first two MAXI GSC energy bands (see Figure \ref{orbital_hardness}). The HR values during the eclipse contain huge error bars; thus, we remove them from the plot. In general, the HR varies between 0.1 and 0.4 during the noneclipse phases. We do not observe clear spectral variability corresponding to the sinusoidal profile. A hardening feature is marginally seen at $\phi_{\rm{orb}}\sim$0.7--0.8 during the superorbital transition states, although the uncertainty is large. This HR change is not seen in the averaged orbital profile obtained with the ASM \citep[see, e.g., Figure 3 of][]{Trowbridge2007} and could be caused by both the insufficient photon statistic of ASM and the superorbital dependence of the dip. 

\subsection{Superorbital and Orbital Profiles in the Optical Band} \label{optical_section}

\begin{figure}
\includegraphics[width=0.47\textwidth]{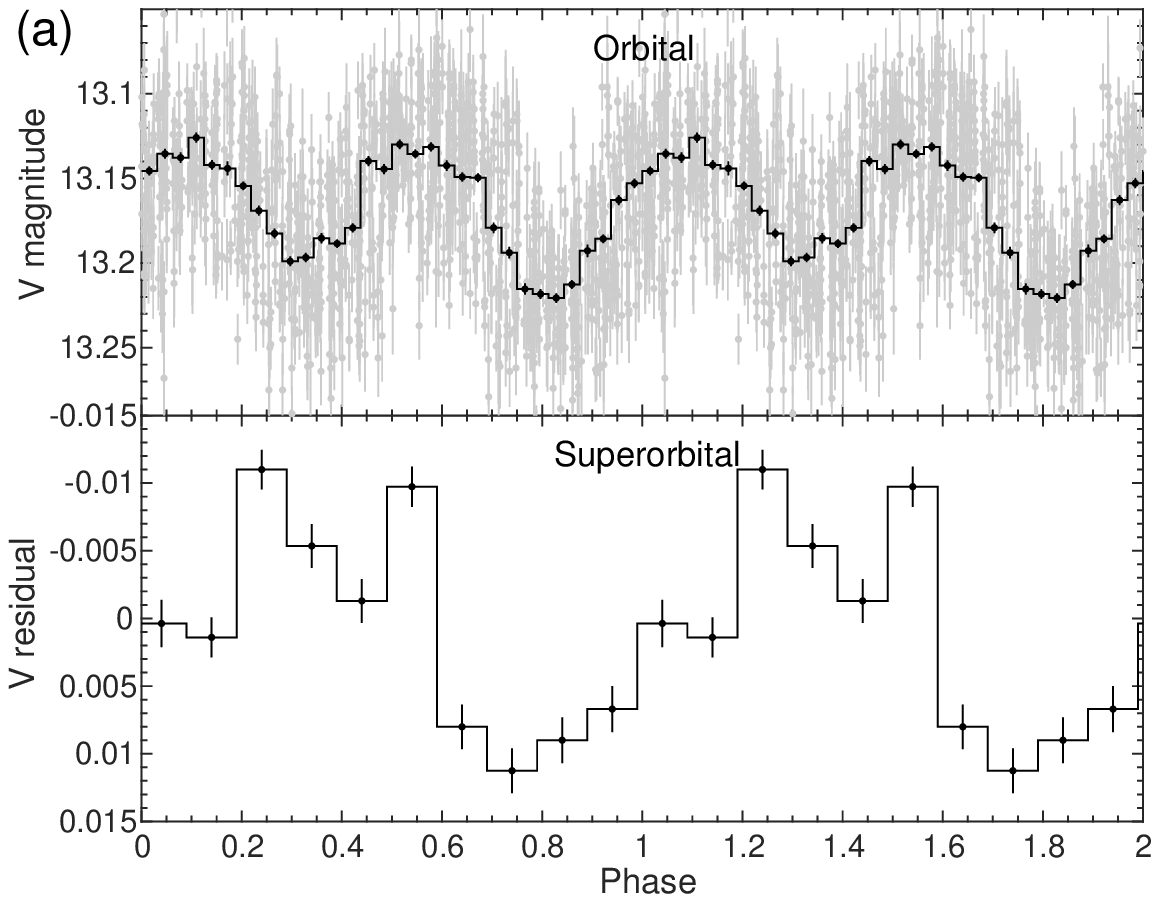} 
\includegraphics[width=0.47\textwidth]{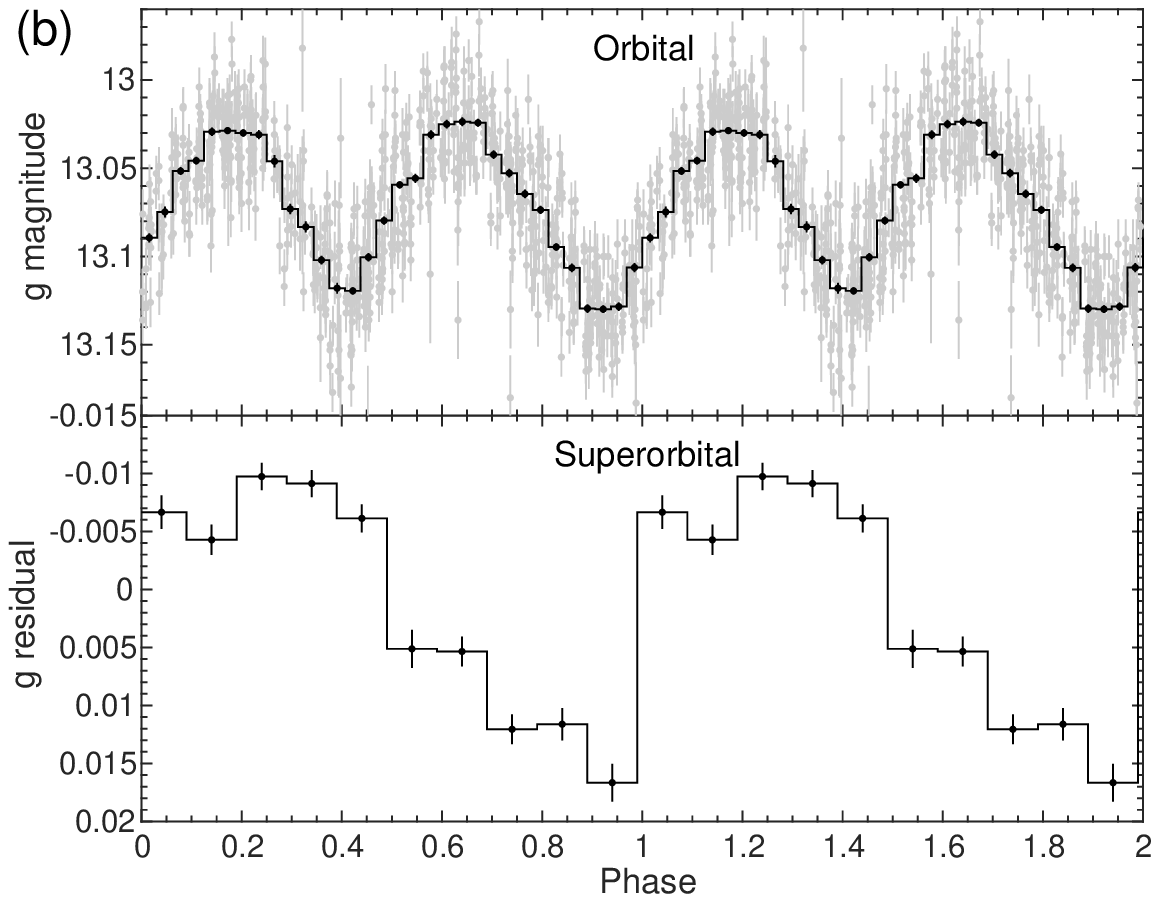} 
\caption{The ASAS-SN folded light curves in the (a) $V$ and (b) $g$ band. The gray points in the upper panel of each figure are magnitude measurements versus orbital phase, where the histogram shows the average orbital profile with 32 bins. The lower panel shows the superorbital profile of the residual after removing the orbital modulation.   \label{fold_asassn}}
\end{figure}

The orbital modulation of SMC X-1 can be seen in the optical band with the OGLE survey data \citep{Udalski2003, CoeAO2013}. Moreover, the superorbital modulation is also weakly determined. We try to use the archival ASAS-SN data to see if the optical modulation can be independently obtained with a different survey program. We fold the light curve according to the orbital ephemeris and find that the orbital modulation can be seen in both the $V$ and $g$ bands (Figure \ref{fold_asassn}). The optical profile is double-peaked, likely caused by the ellipsoidal modulation of a Roche lobe filling companion star. Similar to the result obtained in \citet{CoeAO2013}, the flux minimum at $\phi_{\rm{orb}}=0.5$ is brighter than that at $\phi_{\rm{orb}}=0$, which is a feature of X-ray heating of the companion's surface \citep{RawlsOM2011}.

We then attempt to fold the light curves according to the superorbital phase defined with the HHT but fail to obtain a significant modulation profile. This could be caused by the contamination of the strong orbital modulation because only $\sim$3\% of the optical emission is expected to modulate at the superorbital period \citep{CoeAO2013}. Therefore, we create a template by smoothing the orbital profile. Then, we get the residual magnitude of each observation by subtracting the template from the observed magnitude. The residual has an RMS scattering of 0.04 mag in the $V$ band and 0.02 mag in the $g$ band. We fold the residual light curve according to the superorbital phase and obtain superorbital profiles (Figure \ref{fold_asassn}). Both the $V$- and $g$-band profiles show a clear modulation with a valley at $\phi_{\rm{sup}}\approx$0.7--1.  The peak-to-peak amplitudes of the $V$- and $g$-band superorbital profiles are $\sim$0.02 and $\sim$0.025 mag, respectively. We further obtain the superorbital profile using the data collected in MJD 58002--58385, where SMC X-1 is simultaneously observed with both the $V$ and $g$ bands although the observational cadences between these two bands are different. The result is fully consistent with that observed from the entire data set.


\section{Discussion}\label{discussion}
\subsection{Implication of Superorbital Period Excursion}
For the first time, we extend the study of the superorbital modulation to $\sim$7 yr after the end of the \rxte\ mission. A new major superorbital excursion event is observed independently with the \swift\ BAT and MAXI GSC near $\sim$MJD 57100. This suggests that the superorbital excursion event could have a (quasi)periodicity of $\sim$3100-3200 days, which confirms the prediction made by \citet{Hu2011}. If this is true, a period excursion is expected to occur at $\sim$MJD 47600. Unfortunately, the \emph{BATSE} observation does not cover this epoch, but a possible fluctuation in $f_{\rm{sup}}$ can be seen before $\sim$MJD 49000 \citep{ClarksonCC2003a}. 

The instantaneous frequencies obtained with the \rxte\ ASM, \swift\ BAT, and MAXI GSC have moderate differences near the boundaries of individual data sets. The discrepancies are caused by the end effect of the EMD that leads to low robustness in the instantaneous frequency measurements. Except for the boundaries, the instantaneous frequencies observed by different satellites are generally consistent with each other. This indicates that not only the excursion events described in the previous paragraph but also the short-term variabilities are intrinsic features of SMC X-1. These minor fluctuations have been noticed in \citep{Hu2011} solely using the \rxte\ ASM. For example, the superorbital frequency showed a double-peaked feature near $\sim$54200--54500. This event is clearly seen regardless of the analysis techniques (WWZ and HHT) and the observation instrument (\rxte\ ASM and \swift\ BAT). Similar events are also seen in $\sim$MJD 52200--52500 and $\sim$MJD 55200--55500. 

A possible hint of a weak correlation between the superorbital modulation period and modulation amplitude is observed \citep{Hu2011}. We confirm that during the excursion epoch, the superorbital profile has a low amplitude, which is dominated by the increase of low-state flux.

The current leading model of the superorbital modulation is the occultation by a precessing and warped disk \citep{Wijers1999, OgilvieD2001}. Such a warp could be driven by the perturbation from the strong radiation \citep{Petterson1977, Pringle1996, Wijers1999}. The stability of the warp is characterized by the binary separation ($r_b$ in units of $GM_{\rm{NS}}/c^2$) and $q$. A large $r_b/q$ value suggests a highly unstable warp mode \citep{OgilvieD2001}. Both stable mode 0 and unstable mode 1 precession could be contained in SMC X-1 and hence cause the drift of the superorbital modulation period \citep{ClarksonCC2003a, Charles2008}. 

The disk warp model predicts a connection between the disk luminosity and the superorbital modulation period as
\begin{equation}\label{equation_p_lum}
P_{\rm{sup}}\sim\frac{\Sigma_d R_d^{3/2}}{L_d},
\end{equation}
where $\Sigma_d$ is the surface density of the disk, $R_d$ is a characteristic radius of the disk, and $L_d$ is the strength of the radiation field from the disk \citep{Wijers1999, Still2004}.  However, the decreasing trend of the high-state flux between the first and the second excursion events does not result in an increasing trend of the superorbital cycle length. This implies that the mass accretion rate is not strongly associated with the superorbital period.

Owing to the low inclination angle \citep[$i=65^{\circ}$;][]{BakerNQ2005} of SMC X-1, the increment in the low-state flux during the excursion epochs can be interpreted as the decrease of warp because the emission near the neutron star may not be fully obscured during the superorbital low state. The warp angle is expected to vary with time quasiperiodically or aperiodically, although its connection to the precession period remains unclear \citep{Pringle1997}. Furthermore, the gradual change of the hardness ratio during the superorbital transition state suggests that the atmosphere of the disk is not fully opaque. Variability in the warp angle could change the obscuring percentage of the X-ray emission and the optical depth along the line of sight during the superorbital transition and low states. This results in the variability of the low-state flux as seen in our analysis.


Utilizing the ASAS-SN data, we independently obtain the superorbital modulation profile in the optical band (see Figure \ref{fold_asassn}). The peak-to-peak amplitude of the superorbital modulation roughly corresponds to a $2.3$\% flux difference, in the same order as that obtained by \citet{CoeAO2013} by using the OGLE data. The optical profile is in phase with the X-ray superorbital profile, implying that the superorbital modulation in the optical band is from the same source as the superorbital modulation in the X-ray band. The optical emission from the disk is believed to be the reprocessed X-ray \citep{Howarth1982}. A global warp of the disk with a maximum disk warp angle at the edge and with a finite tilt angle could interpret the observational result \citep{FoulkesHM2010}. During the superorbital low state, the X-ray-illuminated side of the disk is partially obscured by the backlight side of the warp.  The presence of the superorbital phase-dependent pre-eclipse dip also suggests a deviation of the outermost disk from the orbital plane. During the superorbital high state, the near side of the outer disk is far from our line of sight; hence, the bulge could not pass through our line of sight. Given that the accretion disk contributes $\sim$5\% of the optical emission \citep{Howarth1982}, roughly 40\% of the disk emission in the optical band is obscured during the superorbital low state.

Alternatively, the companion star could be heated by the neutron star, and the warped disk could cause a shadow on the companion. This could drive the superorbital modulation in the optical band and change the orbital profile along with the superorbital phase \citep[see, e.g.,][]{DeeterCG1976, GerendB1976}. However, the superorbital modulation in the optical band would be more likely be out of phase with the X-ray modulation. This is not consistent with the observation but could be an origin that reduces the observed optical modulation amplitude.

\subsection{Spin-Superorbital Connection}

The high time resolution and the large effective area of the MAXI GSC allow us to track the spin period of SMC X-1 and refine the orbital parameters. We find that $a\sin i/c$, $e$, and $\Omega$ are not well constrained compared to several pointed observations \citep[see, e.g.,][]{RaichurP2010}. Moreover, their values are not consistent with each other from different measurements \citep{Inam2010, RaichurP2010}. This could be caused by the variability of the pulse profile \citep{RaichurP2010}. Previous observations suggest that the pulse profile of SMC X-1 only varies dramatically in the soft X-ray band \citep{Neilsen2004, Hickox2005} but recent \nustar\ observations suggest that the profile is likely superorbital phase-dependent in the hard X-ray \citep{PikeHB2019}. Our study shows that the averaged pulse profile varies between superorbital cycles in the 2--20\,keV band (see Figure \ref{2dsearch_maxi}). This could be originated not only from the intrinsic variability of the pulse profile but also from the variability of $\dot{\nu}$ within one superorbital cycle. In a few superorbital segments that result in a high $Z_2^2$ values, we find that $\nu$ has a clear drift with a variable $\dot{\nu}$ rate. This could distort the pulse profile if we assume a fixed $\dot{\nu}$ within a superorbital cycle. Nevertheless, these two effects prevent us from measuring a precise orbital eccentricity and the angle of periastron.

It is suggested that the spin period evolution of SMC X-1 could have a connection to the variability of the superorbital modulation \citep{DageCC2018}. We do not observe a clear positive correlation between superorbital frequency and the spin frequency residual with the MAXI GSC and hence suggest that the apparent correlation observed with \rxte\ could be a coincidence.

The spin-up rate of an accreting pulsar is dominated by the change of mass accretion rate. The spin-up torque ($N$) is expressed as
\begin{equation}
N\approx\dot{M}\sqrt{GM_{\rm{NS}}r_m},
\end{equation}
where $\dot{M}$ is the mass accretion rate, $M_{\rm{NS}}$ is the mass of the neutron star, and $r_m$ is the magnetospheric radius. The value of $r_m$ also depends on $\dot{M}$ as
\begin{equation}
r_m\propto\left( \frac{\mu^4}{M_{\textrm{NS}}\dot{M}^2} \right)^\frac{1}{7},
\end{equation}
where $\mu$ is the magnetic moment of the neutron star \citep{PringleR1972, GhoshL1979, BildstenCC1997}. When $r_m$ is smaller than the corotating radius, the neutron star spins up with a rate positively correlated with the mass accretion rate. If this is the case for SMC X-1, a correlation between the spin-up rate and the high-state flux is expected. We do not observe a clear flux increase during the first superorbital excursion; hence, the spin period change is likely not caused by the change of the mass accretion rate. The X-ray flux between the first excursion epoch and the second excursion epochs likely has a decreasing trend (Figure \ref{high_low_flux}). This could result in a decrease in the spin-up trend and partly cause the overestimate of the frequency after MJD 55000 by fitting spin period measurements in MJD 50800--53000 with a straight line.

On the other hand, the angular momenta of the materials on a warped disk also depend on the warp angle. If the excursion of the superorbital period is caused by the instability of the warp, a connection between the spin-up rate change and the superorbital modulation period is expected. This could be tested using MAXI GSC data because the high-state flux during MAXI observations is relatively stable. We indeed observe a hint of anticorrelation that is inconsistent with the behavior observed with \rxte\ near the first superorbital excursion event (see Section \ref{2dsearch_section}).


However, the apparent anticorrelation could be a coincidence. The superorbital period excursion could permanently change the disk configuration, and the warp inclination may not exactly return to the pre-excursion value. This could cause a sudden change of accretion torque as suggested in \citet{DageCC2018} although a time lag between the spin period change and superorbital period excursion is possible. If this is true, the third superorbital period excursion event increases the accretion torque. This is different from that in the first superorbital period excursion event. Moreover, the torque could also be changed during the minor excursion events of the superorbital period. Several minor superorbital excursion events coincide with the local minima of the spin period residual (see Figure \ref{spin_residual}), but a few peaks of the superorbital frequency are likely with the local maxima of the frequency residual.  Therefore, the apparent anticorrelation during the MAXI GSC observation could be a biased effect if we observe more minor excursion events associated with positive torque changes than those events associated with negative changes.

The spin period evolution of SMC X-1 is complicated and could be associated with the change of the mass accretion rate and the variability of the warp angle. Therefore, we could not obtain a global correlation between the superorbital and spin frequencies. Our result suggests that the excursion in the superorbital modulation period could change the accretion torque, but the size and the direction of change remain unclear. Therefore, we could not entirely exclude the possibility that the spin period evolution is unassociated with the change of the superorbital period. Monitoring the spin and superorbital behaviors of SMC X-1 would be helpful to probe the spin-superorbital connection.

\subsection{Comparison to Her X-1}
Similar to SMC X-1, Her X-1 is a famous X-ray binary that shows fruitful irregularities in the superorbital modulation amplitude and pulse profile and a connection between spin and superorbital modulation. Her X-1 shows a positive correlation between the superorbital period and ASM flux at main-on, not consistent with Equation \ref{equation_p_lum} \citep{Still2004, Leahy2010}.   Her X-1 is argued to have a luminosity higher than the threshold value that causes the self-occultation of the disk. Strong emission from the neutron star effectively reduces the radiation pressure and increases the superorbital period \citep{Still2004}. Different from Her X-1, our result shows that the correlation between modulation amplitude and superorbital cycle length is dominated by the low-state flux, which is not directly associated with the mass accretion rate and the disk luminosity. The direct observational evidence is that the pulse profile of Her X-1 at all energies varies with the superorbital cycle \citep{ScottLW2000, StaubertKV2013}, while for SMC X-1, the pulse profile only varies at soft energies \citep{Hickox2005}, although recent \nustar\ observations show that the pulse profile of SMC X-1 is also moderately variable in the hard X-ray \citep{PikeHB2019}.  

The spin period of Her X-1 does not monotonically increase with time. It occasionally spins down when the main-on flux is low. The pulsar in Her X-1 significantly spins down during the anomalous low states, in which the superorbital modulation cannot be seen \citep[see, e.g.,][and the \fermi\ GBM pulsar monitoring program\footnote{\url{https://gammaray.msfc.nasa.gov/gbm/science/pulsars.html}}]{StaubertKP2009}. It is suggested that the warp angle of the innermost accretion disk could be higher than 90$^{\circ}$ and result in a spin-down torque during the anomalous low state \citep{ParmarO1999}. The orbital profile of Her X-1 in the optical band is found to vary over the superorbital phase \citep{DeeterCG1976, GerendB1976}. However, the orbital profile at any specific superorbital phase, as well as the superorbital profile in the optical band, remain consistent between the normal state and the anomalous low state \citep{JuruaCS2011}. This suggests that the change of the disk warp is very slight between these two states. Another explanation is that the decrease of the mass accretion rate enlarges the magnetospheric radius to be larger than the corotating radius \citep{Leahy2010}. The accretion is then prohibited due to the centrifugal barrier and causes the pulsar spin down. We do not observe a significant spin-down torque in SMC X-1; hence, we suggest that SMC X-1 remains far from the equilibrium and the change in warp angle is not extremely strong.

\subsection{Precession of a Ring Tube}\label{warp_disk}
The superorbital profile of SMC X-1 could also be modeled by the tidal-induced precession of a tilted disk, where the X-ray is quasiperiodically obscured by the ring tube on the disk \citep{Inoue2012, Inoue2019}.  This model can reproduce the light-curve shape of SMC X-1, LMC X-4, and Her X-1. The geometrically thick ring is formed at the outermost part of the accretion disk with a radius of
\begin{equation}
\frac{R_{\rm{ring}}}{a}=\left[ 2\frac{(1+q)^{1/2}}{q} \frac{P_{\rm{orb}}}{P_{\rm{sup}}}\frac{1}{\cos\theta} \right]^{2/3},
\end{equation}
where $R_{\rm{ring}}$ is the radius of the ring, and $\theta$ is the tilt angle of the ring axis against the precession axis. For SMC X-1, this value is $\sim$0.13. In this model, the warp of the disk is negligibly small \citep{Inoue2012}. The variability of the low-state flux varies due to the change of the opacity of the ring tube, the scale height of the tube, and the tilt angle of the disk. The scale height $x_0$ of the ring tube can be described as
\begin{equation}\label{equation_scale_height}
x_0=\left( \frac{2kT}{m_{\textrm{H}}GM_{\textrm{NS}}} \right)^{\frac{1}{2}}R_{\rm{ring}}^{\frac{3}{2}},
\end{equation}
where $T$ is the disk temperature, $m_{\textrm{H}}$ is the mass of a hydrogen atom, $k$ is the Stefan-Boltzmann constant and $G$ is the gravitational constant. The relation between $T$ and the ionization parameter $\xi$ is $T\propto\xi^\eta$ where $\eta$ is a logarithmic slope. Assuming a constant X-ray luminosity $L_X$, this relation can be written as
\begin{equation}
T\propto(n_0R_{\rm{ring}}^2)^{-\eta},
\end{equation}
where $n_0$ is the gaseous number density at the tube center. According to the mass conservation $n_0x_0^2R_{\rm{ring}}=\rm{constant}$, we can derive
\begin{equation}
x_0\propto R_{\rm{ring}}^{(3+\eta)/2(1-\eta)}
\end{equation}
Moreover, the optical depth $\tau_0$ of the ring tube is evaluated as $\tau_0=\sigma_{\textrm{T}}n_0x_0$, where  $\sigma_{\textrm{T}}$ is the Thompson cross section. We can then calculate the relationship between $\tau_0$ and $R$ as
\begin{equation}
\tau_0\propto R_{\rm{ring}}^{(1+3\eta)/2(1-\eta)}\propto P_{\rm{sup}}^{-(1+3\eta)/3(1-\eta)}.
\end{equation}
As long as $\eta>1$, the optical depth and the scale height of the tube decrease as $P_{\rm{sup}}$ decreases. This increases the low-state flux and reduces the superorbital amplitude. However, the tilt angle increases as $R_{\rm{ring}}$ by $\delta R_{\rm{ring}}/R_{\rm{ring}}\propto \tan\theta\delta\theta$ where $\theta$ is the tilt angle of the disk \citep{Inoue2012}. This increases the obscuring/absorption ratio of the X-ray emission and enhances the observed amplitude. The viewing geometry has to be strongly restricted to reproduce the observed anticorrelation.

The superorbital modulation in the optical band is expected if the emission from the ring is mainly in the optical band. Finally, the connection between the evolution of the spin period and the superorbital modulation could also be interpreted, since the tilt angle of the disk changes with the superorbital modulation period.

\section{Summary}\label{summary}
In this research, we track the evolution of the superorbital period in SMC X-1 with the \rxte\ ASM, \swift\ BAT, and MAXI GSC with a baseline longer than 23 yr. With the high-quality MAXI GSC data, we also study the HR variability, along with the orbital and superorbital phases, and track the spin period evolution. Our achievements and conclusions are as follows.

(i) Three major superorbital excursion events are identified at $\sim$MJD 50800, 54000, and 57100. Several minor excursions are seen during the regular epochs between major excursion events. We suggest that the excursion event is likely recurrent and possible (quasi)periodic. 

(ii) We observe a gradual change of the HR during the superorbital transition state, implying a possible absorption. This could be interpreted if the atmosphere of the warped disk is not entirely opaque.

(iii) From the correlation analysis, we find that the low-state flux increases during the excursion epoch. In contrast, the high-state flux shows less obvious variation, except for the first excursion. This indicates that the X-ray emission from the central region is not fully obscured/absorbed during the excursion epoch.

(iv) A gradual decrease in the high-state flux is found during MJD 51500 and 53500, where no significant trend in the superorbital modulation period is found. We suggest that the change in the superorbital modulation period is not dominated by the mass accretion rate.

(v) We successfully track the spin period evolution with a sequence of two-dimensional period searches on MAXI GSC data. We do not observe a positive correlation between the superorbital frequency and the spin period evolution. On the other hand, a possible weak anticorrelation is observed. This could be interpreted if the major/minor excursion events result in a change of the accretion torque. 

(vi) We refine the orbital ephemeris according to the spin period measurement with the MAXI GSC. 

(vii) An increase in the HR at $\phi_{\rm{orb}}=0.7$--$0.8$ is marginally seen during the superorbital descending state. This supports the idea that SMC X-1 is a dipping HMXB.

(viii) We reconfirm the superorbital modulation in the optical band with ASAS-SN data. The superorbital profile in the optical band is in phase with the X-ray one.

(ix) The ring tube on a precessed and tilted accretion disk provides an alternative interpretation for the superorbital modulation. The observed phenomena could be explained by the change of tilt angle, opacity, and scale height of the ring tube.

\acknowledgments
We thank Prof.~Hajime Inoue for useful discussions. We appreciate valuable comments from the referee to improve this paper. This work made use of data provided by the ASM/\emph{RXTE} teams at MIT and the \rxte\ SOF and GOF at NASA's GSFC, the \swift\ BAT data provided by the hard X-ray transient monitor \citep{KrimmHC2013}, and the MAXI data provided by RIKEN, JAXA, and the MAXI team. This work also made use \swift\ XRT data supplied by the UK Swift Science Data Centre at the University of Leicester and \chandra\ data obtained from the \chandra\ Data Archive. We also made use of observations made with \xmm, which is an ESA science mission with instruments and contributions directly funded by the ESA member states and NASA. C.-P.H. acknowledges support from the Japan Society for the Promotion of Science (JSPS; ID: P18318).

\facilities{\rxte\ (ASM), \swift\ (XRT, BAT), MAXI (GSC), \emph{CXO} (HRC), \emph{XMM} (PN)}
\software{CIAO \citep{FruscioneMA2006}, \xmm\ SAS, HEAsoft}


\begin{thebibliography}{}
\expandafter\ifx\csname natexlab\endcsname\relax\def\natexlab#1{#1}\fi
\providecommand{\url}[1]{\href{#1}{#1}}

\bibitem[{{Barthelmy} {et~al.}(2005){Barthelmy}, {Barbier}, {Cummings},
  {Fenimore}, {Gehrels}, {Hullinger}, {Krimm}, {Markwardt}, {Palmer},
  {Parsons}, {Sato}, {Suzuki}, {Takahashi}, {Tashiro}, \&
  {Tueller}}]{BarthelmyBC2005}
{Barthelmy}, S.~D., {Barbier}, L.~M., {Cummings}, J.~R., {et~al.} 2005, \ssr,
  120, 143

\bibitem[{{Bildsten} {et~al.}(1997){Bildsten}, {Chakrabarty}, {Chiu}, {Finger},
  {Koh}, {Nelson}, {Prince}, {Rubin}, {Scott}, {Stollberg}, {Vaughan},
  {Wilson}, \& {Wilson}}]{BildstenCC1997}
{Bildsten}, L., {Chakrabarty}, D., {Chiu}, J., {et~al.} 1997, \apjs, 113, 367

\bibitem[{{Buccheri} {et~al.}(1983){Buccheri}, {Bennett}, {Bignami}, {Bloemen},
  {Boriakoff}, {Caraveo}, {Hermsen}, {Kanbach}, {Manchester}, {Masnou},
  {Mayer-Hasselwander}, {{\"O}zel}, {Paul}, {Sacco}, {Scarsi}, \&
  {Strong}}]{Buccheri1983}
{Buccheri}, R., {Bennett}, K., {Bignami}, G.~F., {et~al.} 1983, \aap, 128, 245

\bibitem[{{Charles} {et~al.}(2008){Charles}, {Clarkson}, {Cornelisse}, \&
  {Shih}}]{Charles2008}
{Charles}, P., {Clarkson}, W., {Cornelisse}, R., \& {Shih}, C. 2008, New
  Astronomy Reviews, 51, 768

\bibitem[{{Clarkson} {et~al.}(2003){Clarkson}, {Charles}, {Coe}, {Laycock},
  {Tout}, \& {Wilson}}]{ClarksonCC2003a}
{Clarkson}, W.~I., {Charles}, P.~A., {Coe}, M.~J., {et~al.} 2003, \mnras, 339,
  447

\bibitem[{{Cleveland}(1979)}]{Cleveland1979}
{Cleveland}, W.~S. 1979, Journal of the American Statistical Association, 74,
  829.

\bibitem[{{Cleveland} \& {Devlin}(1988)}]{ClevelandD1988}
{Cleveland}, W.~S., \& {Devlin}, S.~J. 1988, Journal of the American
  Statistical Association, 83, 596.

\bibitem[{{Coe} {et~al.}(2013){Coe}, {Angus}, {Orosz}, \&
  {Udalski}}]{CoeAO2013}
{Coe}, M.~J., {Angus}, R., {Orosz}, J.~A., \& {Udalski}, A. 2013, \mnras, 433,
  746

\bibitem[{{Coe} {et~al.}(1981){Coe}, {Burnell}, {Engel}, {Evans}, \&
  {Quenby}}]{CoeBE1981}
{Coe}, M.~J., {Burnell}, S.~J.~B., {Engel}, A.~R., {Evans}, A.~J., \& {Quenby},
  J.~J. 1981, \mnras, 197, 247

\bibitem[{{Corbet}(1986)}]{Corbet1986}
{Corbet}, R.~H.~D. 1986, \mnras, 220, 1047

\bibitem[{{Corbet} \& {Krimm}(2013)}]{CorbetK2013}
{Corbet}, R.~H.~D., \& {Krimm}, H.~A. 2013, \apj, 778, 45

\bibitem[{{Dage} {et~al.}(2019){Dage}, {Clarkson}, {Charles}, {Laycock}, \&
  {Shih}}]{DageCC2018}
{Dage}, K.~C., {Clarkson}, W.~I., {Charles}, P.~A., {Laycock}, S.~G.~T., \&
  {Shih}, I.-C. 2019, \mnras, 482, 337

\bibitem[{{Deeter} {et~al.}(1976){Deeter}, {Crosa}, {Gerend}, \&
  {Boynton}}]{DeeterCG1976}
{Deeter}, J., {Crosa}, L., {Gerend}, D., \& {Boynton}, P.~E. 1976, \apj, 206,
  861

\bibitem[{Efron(1979)}]{Efron1979}
Efron, B. 1979, The Annals of Statistics, 7, 1.

\bibitem[{{Falanga} {et~al.}(2015){Falanga}, {Bozzo}, {Lutovinov},
  {Bonnet-Bidaud}, {Fetisova}, \& {Puls}}]{Falanga2015}
{Falanga}, M., {Bozzo}, E., {Lutovinov}, A., {et~al.} 2015, \aap, 577, A130

\bibitem[{{Foster}(1996)}]{Foster1996}
{Foster}, G. 1996, \aj, 112, 1709

\bibitem[{{Foulkes} {et~al.}(2010){Foulkes}, {Haswell}, \&
  {Murray}}]{FoulkesHM2010}
{Foulkes}, S.~B., {Haswell}, C.~A., \& {Murray}, J.~R. 2010, \mnras, 401, 1275

\bibitem[{{Fruscione} {et~al.}(2006){Fruscione}, {McDowell}, {Allen},
  {Brickhouse}, {Burke}, {Davis}, {Durham}, {Elvis}, {Galle}, {Harris},
  {Huenemoerder}, {Houck}, {Ishibashi}, {Karovska}, {Nicastro}, {Noble},
  {Nowak}, {Primini}, {Siemiginowska}, {Smith}, \& {Wise}}]{FruscioneMA2006}
{Fruscione}, A., {McDowell}, J.~C., {Allen}, G.~E., {et~al.} 2006, \procspie,
  6270, 62701V

\bibitem[{{Gerend} \& {Boynton}(1976)}]{GerendB1976}
{Gerend}, D., \& {Boynton}, P.~E. 1976, \apj, 209, 562

\bibitem[{{Ghosh} \& {Lamb}(1979)}]{GhoshL1979}
{Ghosh}, P., \& {Lamb}, F.~K. 1979, \apj, 234, 296

\bibitem[{Grossmann \& Morlet(1984)}]{GrossmannM1984}
Grossmann, A., \& Morlet, J. 1984, SIAM Journal on Mathematical Analysis, 15,
  723.

\bibitem[{{Gruber} \& {Rothschild}(1984)}]{Gruber1984}
{Gruber}, D.~E., \& {Rothschild}, R.~E. 1984, \apj, 283, 546

\bibitem[{{Henry} \& {Schreier}(1977)}]{HenryS1977}
{Henry}, P., \& {Schreier}, E. 1977, \apjl, 212, L13

\bibitem[{{Hickox} \& {Vrtilek}(2005)}]{Hickox2005}
{Hickox}, R.~C., \& {Vrtilek}, S.~D. 2005, \apj, 633, 1064

\bibitem[{Holland \& Welsch(1977)}]{HollandW1977}
Holland, P.~W., \& Welsch, R.~E. 1977, Communications in Statistics - Theory
  and Methods, 6, 813.

\bibitem[{{Howarth}(1982)}]{Howarth1982}
{Howarth}, I.~D. 1982, \mnras, 198, 289

\bibitem[{{Hu} {et~al.}(2017){Hu}, {Chou}, {Ng}, {Lin}, \& {Yen}}]{HuCN2017}
{Hu}, C.-P., {Chou}, Y., {Ng}, C.-Y., {Lin}, L.~C.-C., \& {Yen}, D.~C.-C. 2017,
  \apj, 844, 16

\bibitem[{{Hu} {et~al.}(2011){Hu}, {Chou}, {Wu}, {Yang}, \& {Su}}]{Hu2011}
{Hu}, C.-P., {Chou}, Y., {Wu}, M.-C., {Yang}, T.-C., \& {Su}, Y.-H. 2011, \apj,
  740, 67

\bibitem[{{Hu} {et~al.}(2013){Hu}, {Chou}, {Yang}, \& {Su}}]{Hu2013}
{Hu}, C.-P., {Chou}, Y., {Yang}, T.-C., \& {Su}, Y.-H. 2013, \apj, 773, 58

\bibitem[{{Hu} {et~al.}(2014){Hu}, {Chou}, {Yang}, \& {Su}}]{Hu2014}
---. 2014, \apj, 788, 31

\bibitem[{{Huang} {et~al.}(1998){Huang}, {Shen}, {Long}, {Wu}, {Shih}, {Zheng},
  {Yen}, {Tung}, \& {Liu}}]{Huang1998}
{Huang}, N.~E., {Shen}, Z., {Long}, S.~R., {et~al.} 1998, Royal Society of
  London Proceedings Series A, 454, 903

\bibitem[{{Huber}(1981)}]{Huber1981}
{Huber}, P.~J. 1981, {Robust statistics}, 1981, Wiley Series in Probability and Mathematical Statistics, New York: Wiley

\bibitem[{\.{I}nam {et~al.}(2010)\.{I}nam, Baykal, \& Beklen}]{Inam2010}
\.{I}nam, S.~c., Baykal, A., \& Beklen, E. 2010, \mnras, 403, 378

\bibitem[{{Inoue}(2012)}]{Inoue2012}
{Inoue}, H. 2012, \pasj, 64, 40

\bibitem[{{Inoue}(2019)}]{Inoue2019}
---. 2019, \pasj, 71, 36

\bibitem[{{Jenke} {et~al.}(2012){Jenke}, {Finger}, {Wilson-Hodge}, \&
  {Camero-Arranz}}]{JenkeFW2012}
{Jenke}, P.~A., {Finger}, M.~H., {Wilson-Hodge}, C.~A., \& {Camero-Arranz}, A.
  2012, \apj, 759, 124

\bibitem[{{Jurua} {et~al.}(2011){Jurua}, {Charles}, {Still}, \&
  {Meintjes}}]{JuruaCS2011}
{Jurua}, E., {Charles}, P.~A., {Still}, M., \& {Meintjes}, P.~J. 2011, \mnras,
  418, 437

\bibitem[{{Kochanek} {et~al.}(2017){Kochanek}, {Shappee}, {Stanek}, {Holoien},
  {Thompson}, {Prieto}, {Dong}, {Shields}, {Will}, {Britt}, {Perzanowski}, \&
  {Pojma{\'n}ski}}]{KochanekSS2017}
{Kochanek}, C.~S., {Shappee}, B.~J., {Stanek}, K.~Z., {et~al.} 2017, \pasp,
  129, 104502

\bibitem[{{Krimm} {et~al.}(2013){Krimm}, {Holland}, {Corbet}, {Pearlman},
  {Romano}, {Kennea}, {Bloom}, {Barthelmy}, {Baumgartner}, {Cummings},
  {Gehrels}, {Lien}, {Markwardt}, {Palmer}, {Sakamoto}, {Stamatikos}, \&
  {Ukwatta}}]{KrimmHC2013}
{Krimm}, H.~A., {Holland}, S.~T., {Corbet}, R.~H.~D., {et~al.} 2013, \apjs,
  209, 14

\bibitem[{{Leahy} \& {Igna}(2010)}]{Leahy2010}
{Leahy}, D.~A., \& {Igna}, C.~D. 2010, \apj, 713, 318

\bibitem[{{Levine} {et~al.}(1993){Levine}, {Rappaport}, {Deeter}, {Boynton}, \&
  {Nagase}}]{LevineRD1993}
{Levine}, A., {Rappaport}, S., {Deeter}, J.~E., {Boynton}, P.~E., \& {Nagase},
  F. 1993, \apj, 410, 328

\bibitem[{{Levine} {et~al.}(1996){Levine}, {Bradt}, {Cui}, {Jernigan},
  {Morgan}, {Remillard}, {Shirey}, \& {Smith}}]{LevineBC1996}
{Levine}, A.~M., {Bradt}, H., {Cui}, W., {et~al.} 1996, \apjl, 469, L33

\bibitem[{{Levine} {et~al.}(2011){Levine}, {Bradt}, {Chakrabarty}, {Corbet}, \&
  {Harris}}]{LevineBC2011}
{Levine}, A.~M., {Bradt}, H.~V., {Chakrabarty}, D., {Corbet}, R.~H.~D., \&
  {Harris}, R.~J. 2011, \apjs, 196, 6

\bibitem[{{Li} \& {van den Heuvel}(1997)}]{LiV1997}
{Li}, X.-D., \& {van den Heuvel}, E.~P.~J. 1997, \aap, 321, L25

\bibitem[{{Lucke} {et~al.}(1976){Lucke}, {Yentis}, {Friedman}, {Fritz}, \&
  {Shulman}}]{Lucke1976}
{Lucke}, R., {Yentis}, D., {Friedman}, H., {Fritz}, G., \& {Shulman}, S. 1976,
  \apjl, 206, L25

\bibitem[{{Matsuoka} {et~al.}(2009){Matsuoka}, {Kawasaki}, {Ueno}, {Tomida},
  {Kohama}, {Suzuki}, {Adachi}, {Ishikawa}, {Mihara}, {Sugizaki}, {Isobe},
  {Nakagawa}, {Tsunemi}, {Miyata}, {Kawai}, {Kataoka}, {Morii}, {Yoshida},
  {Negoro}, {Nakajima}, {Ueda}, {Chujo}, {Yamaoka}, {Yamazaki}, {Nakahira},
  {You}, {Ishiwata}, {Miyoshi}, {Eguchi}, {Hiroi}, {Katayama}, \&
  {Ebisawa}}]{MatsuokaKU2009}
{Matsuoka}, M., {Kawasaki}, K., {Ueno}, S., {et~al.} 2009, \pasj, 61, 999

\bibitem[{{Mihara} {et~al.}(2011){Mihara}, {Nakajima}, {Sugizaki}, {Serino},
  {Matsuoka}, {Kohama}, {Kawasaki}, {Tomida}, {Ueno}, {Kawai}, {Kataoka},
  {Morii}, {Yoshida}, {Yamaoka}, {Nakahira}, {Negoro}, {Isobe}, {Yamauchi}, \&
  {Sakurai}}]{MiharaNS2011}
{Mihara}, T., {Nakajima}, M., {Sugizaki}, M., {et~al.} 2011, \pasj, 63, S623

\bibitem[{{Neilsen} {et~al.}(2004){Neilsen}, {Hickox}, \&
  {Vrtilek}}]{Neilsen2004}
{Neilsen}, J., {Hickox}, R.~C., \& {Vrtilek}, S.~D. 2004, \apjl, 616, L135

\bibitem[{Nelder \& Mead(1965)}]{NelderM1962}
Nelder, J.~A., \& Mead, R. 1965, The Computer Journal, 7, 308.

\bibitem[{{Ogilvie} \& {Dubus}(2001)}]{OgilvieD2001}
{Ogilvie}, G.~I., \& {Dubus}, G. 2001, \mnras, 320, 485

\bibitem[{{Parmar} {et~al.}(1999){Parmar}, {Oosterbroek}, {dal Fiume},
  {Orlandini}, {Santangelo}, {Segreto}, \& {del Sordo}}]{ParmarO1999}
{Parmar}, A.~N., {Oosterbroek}, T., {dal Fiume}, D., {et~al.} 1999, \aap, 350,
  L5

\bibitem[{{Petterson}(1977)}]{Petterson1977}
{Petterson}, J.~A. 1977, \apj, 218, 783

\bibitem[{{Pike} {et~al.}(2019){Pike}, {Harrison}, {Bachetti}, {Brumback},
  {F{\"u}rst}, {Madsen}, {Pottschmidt}, {Tomsick}, \& {Wilms}}]{PikeHB2019}
{Pike}, S.~N., {Harrison}, F.~A., {Bachetti}, M., {et~al.} 2019, \apj, 875, 144

\bibitem[{{Price} {et~al.}(1971){Price}, {Groves}, {Rodrigues}, {Seward},
  {Swift}, \& {Toor}}]{PriceGR1971}
{Price}, R.~E., {Groves}, D.~J., {Rodrigues}, R.~M., {et~al.} 1971, \apjl, 168,
  L7

\bibitem[{{Pringle}(1996)}]{Pringle1996}
{Pringle}, J.~E. 1996, \mnras, 281, 357

\bibitem[{{Pringle}(1997)}]{Pringle1997}
---. 1997, \mnras, 292, 136

\bibitem[{{Pringle} \& {Rees}(1972)}]{PringleR1972}
{Pringle}, J.~E., \& {Rees}, M.~J. 1972, \aap, 21, 1

\bibitem[{{Raichur} \& {Paul}(2010)}]{RaichurP2010}
{Raichur}, H., \& {Paul}, B. 2010, \mnras, 401, 1532

\bibitem[{{Rawls} {et~al.}(2011){Rawls}, {Orosz}, {McClintock}, {Torres},
  {Bailyn}, \& {Buxton}}]{RawlsOM2011}
{Rawls}, M.~L., {Orosz}, J.~A., {McClintock}, J.~E., {et~al.} 2011, \apj, 730,
  25

\bibitem[{{Reynolds} {et~al.}(1993){Reynolds}, {Hilditch}, {Bell}, \&
  {Hill}}]{Reynolds1993}
{Reynolds}, A.~P., {Hilditch}, R.~W., {Bell}, S.~A., \& {Hill}, G. 1993,
  \mnras, 261, 337

\bibitem[{{Schreier} {et~al.}(1972){Schreier}, {Giacconi}, {Gursky}, {Kellogg},
  \& {Tananbaum}}]{Schreier1972}
{Schreier}, E., {Giacconi}, R., {Gursky}, H., {Kellogg}, E., \& {Tananbaum}, H.
  1972, \apjl, 178, L71

\bibitem[{{Scott} {et~al.}(2000){Scott}, {Leahy}, \& {Wilson}}]{ScottLW2000}
{Scott}, D.~M., {Leahy}, D.~A., \& {Wilson}, R.~B. 2000, \apj, 539, 392

\bibitem[{{Shappee} {et~al.}(2014){Shappee}, {Prieto}, {Grupe}, {Kochanek},
  {Stanek}, {De Rosa}, {Mathur}, {Zu}, {Peterson}, {Pogge}, {Komossa}, {Im},
  {Jencson}, {Holoien}, {Basu}, {Beacom}, {Szczygie{\l}}, {Brimacombe},
  {Adams}, {Campillay}, {Choi}, {Contreras}, {Dietrich}, {Dubberley},
  {Elphick}, {Foale}, {Giustini}, {Gonzalez}, {Hawkins}, {Howell}, {Hsiao},
  {Koss}, {Leighly}, {Morrell}, {Mudd}, {Mullins}, {Nugent}, {Parrent},
  {Phillips}, {Pojmanski}, {Rosing}, {Ross}, {Sand}, {Terndrup}, {Valenti},
  {Walker}, \& {Yoon}}]{ShappeePG2014}
{Shappee}, B.~J., {Prieto}, J.~L., {Grupe}, D., {et~al.} 2014, \apj, 788, 48

\bibitem[{{Staubert} {et~al.}(2009){Staubert}, {Klochkov}, {Postnov},
  {Shakura}, {Wilms}, \& {Rothschild}}]{StaubertKP2009}
{Staubert}, R., {Klochkov}, D., {Postnov}, K., {et~al.} 2009, \aap, 494, 1025

\bibitem[{{Staubert} {et~al.}(2013){Staubert}, {Klochkov}, {Vasco}, {Postnov},
  {Shakura}, {Wilms}, \& {Rothschild}}]{StaubertKV2013}
{Staubert}, R., {Klochkov}, D., {Vasco}, D., {et~al.} 2013, \aap, 550, A110

\bibitem[{{Still} \& {Boyd}(2004)}]{Still2004}
{Still}, M., \& {Boyd}, P. 2004, \apjl, 606, L135

\bibitem[{{Sugizaki} {et~al.}(2011){Sugizaki}, {Mihara}, {Serino}, {Yamamoto},
  {Matsuoka}, {Kohama}, {Tomida}, {Ueno}, {Kawai}, {Morii}, {Sugimori},
  {Nakahira}, {Yamaoka}, {Yoshida}, {Nakajima}, {Negoro}, {Eguchi}, {Isobe},
  {Ueda}, \& {Tsunemi}}]{SugizakiMS2011}
{Sugizaki}, M., {Mihara}, T., {Serino}, M., {et~al.} 2011, \pasj, 63, S635

\bibitem[{{Takagi} {et~al.}(2016){Takagi}, {Mihara}, {Sugizaki}, {Makishima},
  \& {Morii}}]{TakagiMS2016}
{Takagi}, T., {Mihara}, T., {Sugizaki}, M., {Makishima}, K., \& {Morii}, M.
  2016, \pasj, 68, S13

\bibitem[{{Trowbridge} {et~al.}(2007){Trowbridge}, {Nowak}, \&
  {Wilms}}]{Trowbridge2007}
{Trowbridge}, S., {Nowak}, M.~A., \& {Wilms}, J. 2007, \apj, 670, 624

\bibitem[{{Udalski}(2003)}]{Udalski2003}
{Udalski}, A. 2003, \actaa, 53, 291

\bibitem[{{Ulmer} {et~al.}(1973){Ulmer}, {Baity}, {Wheaton}, \&
  {Peterson}}]{UlmerBW1973}
{Ulmer}, M.~P., {Baity}, W.~A., {Wheaton}, W.~A., \& {Peterson}, L.~E. 1973,
  Nature Physical Science, 242, 121
  
\bibitem[{{Val Baker} {et~al.}(2005){Val Baker}, {Norton}, \&
  {Quaintrell}}]{BakerNQ2005}
{Val Baker}, A.~K.~F., {Norton}, A.~J., \& {Quaintrell}, H. 2005, \aap, 441, 685

\bibitem[{{van der Meer} {et~al.}(2007){van der Meer}, {Kaper}, {van Kerkwijk},
  {Heemskerk}, \& {van den Heuvel}}]{vanderMeer2007}
{van der Meer}, A., {Kaper}, L., {van Kerkwijk}, M.~H., {Heemskerk}, M.~H.~M.,
  \& {van den Heuvel}, E.~P.~J. 2007, \aap, 473, 523

\bibitem[{{Wang} {et~al.}(2014){Wang}, {Yeh}, {Young}, {Hu}, \&
  {Lo}}]{WangYY2014}
{Wang}, Y.-H., {Yeh}, C.-H., {Young}, H.-W.~V., {Hu}, K., \& {Lo}, M.-T. 2014,
  Physica A Statistical Mechanics and its Applications, 400, 159

\bibitem[{{Wijers} \& {Pringle}(1999)}]{Wijers1999}
{Wijers}, R.~A.~M.~J., \& {Pringle}, J.~E. 1999, \mnras, 308, 207

\bibitem[{{Wojdowski} {et~al.}(1998){Wojdowski}, {Clark}, {Levine}, {Woo}, \&
  {Zhang}}]{Wojdowski1998}
{Wojdowski}, P., {Clark}, G.~W., {Levine}, A.~M., {Woo}, J.~W., \& {Zhang},
  S.~N. 1998, \apj, 502, 253

\bibitem[{Wu \& Huang(2009)}]{Wu2009}
Wu, Z., \& Huang, N.~E. 2009, AADA, 1, 1.

\bibitem[{{Yatabe} {et~al.}(2018){Yatabe}, {Makishima}, {Mihara}, {Nakajima},
  {Sugizaki}, {Kitamoto}, {Yoshida}, \& {Takagi}}]{YatabeMM2018}
{Yatabe}, F., {Makishima}, K., {Mihara}, T., {et~al.} 2018, \pasj, 70, 89

\bibitem[{{Yeh} {et~al.}(2010){Yeh}, {Shieh}, \& {Huang}}]{YehSH2010}
{Yeh}, J.-R., {Shieh}, J.-S., \& {Huang}, N.~E. 2010, Advances in Adaptive Data
  Analysis, 02, 135.

\end{thebibliography}

\appendix

\section{Measurements of Superorbital Minima}\label{appendix_superorbital}
We list the arrival times of the superorbitl minimum obtained with \rxte\ ASM, \swift\ BAT, and MAXI in Table \ref{all_superorbital_minimum}. The corresponding high- and low-state count rates for the correlation analysis are also included.
\startlongtable
\begin{deluxetable}{l|ccc|ccc|ccc}
\tablecaption{Arrival time of superorbital minima and corresponding high- and low-state count rate measured with ASM, BAT, and MAXI. \label{all_superorbital_minimum}} 
\tablehead{\colhead{Cycle} & \multicolumn{3}{c}{\emph{RXTE} ASM} & \multicolumn{3}{c}{\emph{Swift} BAT} & \multicolumn{3}{c}{MAXI GSC} \\
\colhead{} & \colhead{$T_{\rm{min}}^a$} & \colhead{High$^b$} & \colhead{Low$^c$} & \colhead{$T_{\rm{min}}^a$} & \colhead{High$^b$} & \colhead{Low$^c$} & \colhead{$T_{\rm{min}}^a$} & \colhead{High$^b$} & \colhead{Low$^c$}}
\startdata
1 & 50191.9(8) & \nodata & \nodata & \nodata & \nodata & \nodata & \nodata & \nodata & \nodata \\ 
2 & 50191.9(8) & 2.95(6) & 0.26(8) & \nodata & \nodata & \nodata & \nodata & \nodata & \nodata \\ 
3 & 50252.5(7) & 3.56(5) & 0.27(6) & \nodata & \nodata & \nodata & \nodata & \nodata & \nodata \\ 
4 & 50312.4(4) & 2.96(8) & 0.22(8) & \nodata & \nodata & \nodata & \nodata & \nodata & \nodata \\ 
5 & 50368.2(6) & 2.97(7) & 0.35(13) & \nodata & \nodata & \nodata & \nodata & \nodata & \nodata \\ 
6 & 50418.6(7) & 3.31(6) & 0.40(9) & \nodata & \nodata & \nodata & \nodata & \nodata & \nodata \\ 
7 & 50473.2(6) & 2.29(8) & 0.48(14) & \nodata & \nodata & \nodata & \nodata & \nodata & \nodata \\ 
8 & 50519.5(8) & 2.89(9) & 0.09(10) & \nodata & \nodata & \nodata & \nodata & \nodata & \nodata \\ 
9 & 50572.1(6) & 2.93(9) & 0.41(11) & \nodata & \nodata & \nodata & \nodata & \nodata & \nodata \\ 
10 & 50617.7(6) & 2.54(6) & 0.71(7) & \nodata & \nodata & \nodata & \nodata & \nodata & \nodata \\ 
11 & 50662.5(6) & 2.81(10) & 0.17(10) & \nodata & \nodata & \nodata & \nodata & \nodata & \nodata \\ 
12 & 50705.3(5) & 3.00(17) & 0.57(13) & \nodata & \nodata & \nodata & \nodata & \nodata & \nodata \\ 
13 & 50752.1(7) & 2.79(6) & 0.82(9) & \nodata & \nodata & \nodata & \nodata & \nodata & \nodata \\ 
14 & 50798.0(6) & 2.67(13) & 0.76(16) & \nodata & \nodata & \nodata & \nodata & \nodata & \nodata \\ 
15 & 50839.6(6) & 2.72(12) & -0.14(13) & \nodata & \nodata & \nodata & \nodata & \nodata & \nodata \\ 
16 & 50886.5(6) & 2.66(12) & 0.20(9) & \nodata & \nodata & \nodata & \nodata & \nodata & \nodata \\ 
17 & 50935.5(8) & 2.39(6) & 0.12(12) & \nodata & \nodata & \nodata & \nodata & \nodata & \nodata \\ 
18 & 50981.5(7) & 2.67(8) & 0.56(10) & \nodata & \nodata & \nodata & \nodata & \nodata & \nodata \\ 
19 & 51028.5(7) & 3.17(13) & 0.30(15) & \nodata & \nodata & \nodata & \nodata & \nodata & \nodata \\ 
20 & 51086.4(13) & 2.63(11) & -0.15(14) & \nodata & \nodata & \nodata & \nodata & \nodata & \nodata \\ 
21 & 51141.9(8) & 2.97(10) & 0.40(15) & \nodata & \nodata & \nodata & \nodata & \nodata & \nodata \\ 
22 & 51203.7(9) & 3.26(15) & 0.20(12) & \nodata & \nodata & \nodata & \nodata & \nodata & \nodata \\ 
23 & 51258.3(7) & 2.82(7) & 0.28(6) & \nodata & \nodata & \nodata & \nodata & \nodata & \nodata \\ 
24 & 51314.1(6) & 3.39(7) & 0.35(9) & \nodata & \nodata & \nodata & \nodata & \nodata & \nodata \\ 
25 & 51370.1(7) & 3.33(11) & 0.13(7) & \nodata & \nodata & \nodata & \nodata & \nodata & \nodata \\ 
26 & 51424.5(7) & 3.01(7) & 0.31(9) & \nodata & \nodata & \nodata & \nodata & \nodata & \nodata \\ 
27 & 51485.3(6) & 2.93(6) & 0.01(10) & \nodata & \nodata & \nodata & \nodata & \nodata & \nodata \\ 
28 & 51548.5(4) & 2.96(8) & 0.56(21) & \nodata & \nodata & \nodata & \nodata & \nodata & \nodata \\ 
29 & 51614.7(20) & 2.84(12) & 0.04(12) & \nodata & \nodata & \nodata & \nodata & \nodata & \nodata \\ 
30 & 51673.8(11) & 2.18(7) & 0.13(10) & \nodata & \nodata & \nodata & \nodata & \nodata & \nodata \\ 
31 & 51739.0(7) & 2.78(7) & 0.17(16) & \nodata & \nodata & \nodata & \nodata & \nodata & \nodata \\ 
32 & 51803.1(8) & 2.94(5) & 0.15(11) & \nodata & \nodata & \nodata & \nodata & \nodata & \nodata \\ 
33 & 51859.6(6) & 2.94(5) & 0.29(13) & \nodata & \nodata & \nodata & \nodata & \nodata & \nodata \\ 
34 & 51912.9(7) & 2.72(12) & -0.01(10) & \nodata & \nodata & \nodata & \nodata & \nodata & \nodata \\ 
35 & 51964.7(7) & 2.80(19) & 0.32(8) & \nodata & \nodata & \nodata & \nodata & \nodata & \nodata \\ 
36 & 52017.7(12) & 3.05(5) & 0.17(9) & \nodata & \nodata & \nodata & \nodata & \nodata & \nodata \\ 
37 & 52078.7(5) & 2.95(4) & 0.38(20) & \nodata & \nodata & \nodata & \nodata & \nodata & \nodata \\ 
38 & 52137.3(6) & 2.63(4) & 0.18(8) & \nodata & \nodata & \nodata & \nodata & \nodata & \nodata \\ 
39 & 52193.7(8) & 3.19(9) & 0.17(10) & \nodata & \nodata & \nodata & \nodata & \nodata & \nodata \\ 
40 & 52243.6(7) & 3.07(8) & -0.02(11) & \nodata & \nodata & \nodata & \nodata & \nodata & \nodata \\ 
41 & 52301.0(7) & 2.64(10) & 0.40(12) & \nodata & \nodata & \nodata & \nodata & \nodata & \nodata \\ 
42 & 52364.1(12) & 2.18(10) & 0.32(9) & \nodata & \nodata & \nodata & \nodata & \nodata & \nodata \\ 
43 & 52418.3(7) & 3.04(6) & 0.09(7) & \nodata & \nodata & \nodata & \nodata & \nodata & \nodata \\ 
44 & 52478.5(5) & 2.26(7) & 0.43(14) & \nodata & \nodata & \nodata & \nodata & \nodata & \nodata \\ 
45 & 52545.4(4) & 2.64(6) & 0.12(14) & \nodata & \nodata & \nodata & \nodata & \nodata & \nodata \\ 
46 & 52594.7(6) & 2.58(6) & 0.18(13) & \nodata & \nodata & \nodata & \nodata & \nodata & \nodata \\ 
47 & 52643.3(7) & 2.49(9) & 0.20(13) & \nodata & \nodata & \nodata & \nodata & \nodata & \nodata \\ 
48 & 52695.6(9) & 2.42(10) & 0.07(10) & \nodata & \nodata & \nodata & \nodata & \nodata & \nodata \\ 
49 & 52757.8(6) & 2.79(7) & 0.13(10) & \nodata & \nodata & \nodata & \nodata & \nodata & \nodata \\ 
50 & 52818.5(12) & 2.51(6) & 0.17(9) & \nodata & \nodata & \nodata & \nodata & \nodata & \nodata \\ 
51 & 52871.9(6) & 3.02(10) & 0.00(5) & \nodata & \nodata & \nodata & \nodata & \nodata & \nodata \\ 
52 & 52921.6(6) & 2.47(8) & 0.12(7) & \nodata & \nodata & \nodata & \nodata & \nodata & \nodata \\ 
53 & 52972.1(6) & 2.28(6) & 0.25(17) & \nodata & \nodata & \nodata & \nodata & \nodata & \nodata \\ 
54 & 53025.0(7) & 3.24(10) & 0.45(19) & \nodata & \nodata & \nodata & \nodata & \nodata & \nodata \\ 
55 & 53082.5(16) & 2.11(11) & 0.16(6) & \nodata & \nodata & \nodata & \nodata & \nodata & \nodata \\ 
56 & 53135.4(8) & 2.44(9) & 0.09(8) & \nodata & \nodata & \nodata & \nodata & \nodata & \nodata \\ 
57 & 53190.0(6) & 2.73(10) & -0.04(9) & \nodata & \nodata & \nodata & \nodata & \nodata & \nodata \\ 
58 & 53249.0(7) & 2.73(5) & 0.14(16) & \nodata & \nodata & \nodata & \nodata & \nodata & \nodata \\ 
59 & 53308.4(6) & 2.48(4) & -0.02(10) & \nodata & \nodata & \nodata & \nodata & \nodata & \nodata \\ 
60 & 53365.7(8) & 2.42(11) & 0.05(11) & \nodata & \nodata & \nodata & \nodata & \nodata & \nodata \\ 
61 & 53421.4(7) & 2.44(8) & 0.25(8) & 53484.2(6) & \nodata & \nodata & \nodata & \nodata & \nodata \\ 
62 & 53485.8(12) & 2.37(10) & 0.30(7) & 53547.4(13) & 12.4(4) & 2.0(5) & \nodata & \nodata & \nodata \\ 
63 & 53544.8(13) & 2.18(7) & 0.17(11) & 53609.8(6) & 9.9(7) & 0.5(3) & \nodata & \nodata & \nodata \\ 
64 & 53612.1(6) & 2.40(6) & -0.02(13) & 53666.7(8) & 10.0(2) & 0.8(6) & \nodata & \nodata & \nodata \\ 
65 & 53666.5(7) & 2.35(5) & 0.19(10) & 53722.2(5) & 11.3(3) & 0.3(4) & \nodata & \nodata & \nodata \\ 
66 & 53721.3(5) & 3.02(9) & 0.29(10) & 53775.6(8) & 15.0(5) & 0.3(8) & \nodata & \nodata & \nodata \\ 
67 & 53776.6(7) & 2.14(7) & 0.15(8) & 53829.3(6) & 11.4(4) & 1.6(6) & \nodata & \nodata & \nodata \\ 
68 & 53831.3(9) & 2.00(8) & 0.20(7) & 53878.4(7) & 9.8(4) & 1.0(9) & \nodata & \nodata & \nodata \\ 
69 & 53879.5(8) & 2.11(6) & 0.53(12) & 53925.5(4) & 14.1(3) & 5.2(3) & \nodata & \nodata & \nodata \\ 
70 & 53924.7(10) & 2.05(7) & 0.39(18) & 53969.4(3) & 12.6(2) & 2.4(3) & \nodata & \nodata & \nodata \\ 
71 & 53971.4(7) & 2.21(7) & 0.13(12) & 54016.5(7) & 12.2(2) & 0.6(2) & \nodata & \nodata & \nodata \\ 
72 & 54012.9(9) & 1.98(6) & 0.27(9) & 54058.4(9) & 11.0(2) & 2.7(2) & \nodata & \nodata & \nodata \\ 
73 & 54057.2(8) & 2.05(8) & 0.15(9) & 54101.1(6) & 11.1(3) & 1.7(3) & \nodata & \nodata & \nodata \\ 
74 & 54101.7(6) & 1.95(8) & 0.10(11) & 54150.8(6) & 11.2(2) & 0.9(4) & \nodata & \nodata & \nodata \\ 
75 & 54153.1(6) & 2.34(10) & 0.30(9) & 54203.0(6) & 13.2(3) & 2.2(8) & \nodata & \nodata & \nodata \\ 
76 & 54202.4(11) & 2.02(10) & 0.03(9) & 54258.8(6) & 11.0(3) & 0.3(3) & \nodata & \nodata & \nodata \\ 
77 & 54258.2(9) & 2.40(7) & 0.12(10) & 54317.4(7) & 12.6(4) & 0.8(3) & \nodata & \nodata & \nodata \\ 
78 & 54314.4(9) & 2.50(6) & -0.00(9) & 54373.5(7) & 11.3(3) & 1.2(3) & \nodata & \nodata & \nodata \\ 
79 & 54372.3(5) & 2.88(6) & 0.10(7) & 54436.2(8) & 14.5(2) & 0.9(3) & \nodata & \nodata & \nodata \\ 
80 & 54435.1(6) & 1.96(6) & 0.21(9) & 54494.1(6) & 11.2(3) & 0.1(6) & \nodata & \nodata & \nodata \\ 
81 & 54495.2(6) & 2.50(11) & 0.08(8) & 54545.2(6) & 13.4(5) & -0.2(5) & \nodata & \nodata & \nodata \\ 
82 & 54547.5(8) & 1.86(10) & 0.17(8) & 54596.7(4) & 11.5(2) & 0.1(2) & \nodata & \nodata & \nodata \\ 
83 & 54597.9(5) & 2.60(8) & 0.15(7) & 54652.0(6) & 13.3(3) & 0.8(4) & \nodata & \nodata & \nodata \\ 
84 & 54653.7(6) & 2.47(9) & 0.17(8) & 54711.7(6) & 9.9(2) & 1.1(4) & \nodata & \nodata & \nodata \\ 
85 & 54711.2(7) & 2.15(7) & 0.11(10) & 54767.9(5) & 11.1(2) & 1.4(2) & \nodata & \nodata & \nodata \\ 
86 & 54767.5(7) & 2.01(7) & -0.01(7) & 54815.7(5) & 10.4(2) & 0.3(6) & \nodata & \nodata & \nodata \\ 
87 & 54815.0(5) & 2.07(7) & 0.19(10) & 54868.6(6) & 12.0(3) & 0.3(4) & \nodata & \nodata & \nodata \\ 
88 & 54871.1(8) & 2.43(8) & -0.02(10) & 54932.4(7) & 12.4(2) & 0.3(3) & \nodata & \nodata & \nodata \\ 
89 & 54929.2(7) & 2.28(6) & -0.07(11) & 54995.1(6) & 12.8(3) & 0.4(4) & \nodata & \nodata & \nodata \\ 
90 & 54996.7(9) & 2.66(8) & 0.07(8) & 55055.9(6) & 10.8(3) & 0.8(4) & \nodata & \nodata & \nodata \\ 
91 & 55057.2(6) & 2.36(8) & 0.06(9) & 55112.1(6) & 11.1(3) & 1.0(2) & 55112.8(5) & \nodata & \nodata \\ 
92 & 55113.7(12) & 1.89(7) & 0.11(7) & 55168.6(5) & 11.6(2) & 0.7(3) & 55169.3(4) & 0.151(2) & 0.016(3) \\ 
93 & 55169.6(5) & 2.70(9) & 0.23(13) & 55224.5(7) & 12.1(2) & 1.6(5) & 55225.1(5) & 0.157(3) & 0.005(2) \\ 
94 & 55225.2(7) & 2.44(9) & 0.44(21) & 55282.2(6) & 14.3(3) & 0.3(3) & 55281.8(6) & 0.141(3) & 0.010(3) \\ 
95 & 55282.0(16) & 2.40(12) & 0.18(12) & 55334.7(7) & 11.1(4) & 1.1(4) & 55334.6(4) & 0.158(3) & 0.006(3) \\ 
96 & 55335.3(7) & 2.31(9) & 0.33(12) & 55388.0(6) & 11.7(2) & 0.7(4) & 55388.1(5) & 0.156(3) & 0.008(3) \\ 
97 & 55386.8(8) & \nodata & \nodata & 55448.7(6) & 11.4(4) & 0.9(4) & 55449.5(8) & 0.143(3) & 0.007(3) \\ 
98 & \nodata & \nodata & \nodata & 55507.8(5) & 11.0(2) & 0.8(2) & 55507.4(3) & 0.136(2) & 0.004(3) \\ 
99 & \nodata & \nodata & \nodata & 55567.6(9) & 11.2(3) & 0.4(4) & 55568.0(5) & 0.163(3) & -0.001(3) \\ 
100 & \nodata & \nodata & \nodata & 55627.9(7) & 11.1(4) & 0.1(6) & 55628.1(6) & 0.182(3) & 0.014(4) \\ 
101 & \nodata & \nodata & \nodata & 55684.1(5) & 12.8(3) & 0.7(4) & 55683.7(4) & 0.148(3) & 0.006(3) \\ 
102 & \nodata & \nodata & \nodata & 55732.6(5) & 12.0(4) & 0.7(4) & 55733.9(5) & 0.168(3) & 0.010(3) \\ 
103 & \nodata & \nodata & \nodata & 55788.4(4) & 13.9(4) & -0.0(3) & 55788.6(5) & 0.166(3) & 0.000(3) \\ 
104 & \nodata & \nodata & \nodata & 55841.7(4) & 11.7(2) & 1.2(3) & 55843.6(4) & 0.167(3) & 0.013(4) \\ 
105 & \nodata & \nodata & \nodata & 55896.4(3) & 13.3(2) & 2.1(3) & 55896.2(4) & 0.147(3) & 0.003(3) \\ 
106 & \nodata & \nodata & \nodata & 55950.9(8) & 11.2(2) & 1.2(4) & 55950.5(2) & 0.131(3) & 0.022(3) \\ 
107 & \nodata & \nodata & \nodata & 55998.6(6) & 8.7(3) & 2.2(4) & 55997.9(5) & 0.155(3) & 0.007(6) \\ 
108 & \nodata & \nodata & \nodata & 56049.1(5) & 12.5(4) & 1.2(3) & 56048.7(5) & 0.169(4) & 0.003(4) \\ 
109 & \nodata & \nodata & \nodata & 56105.2(5) & 12.9(4) & 0.1(4) & 56105.0(6) & 0.184(4) & 0.008(4) \\ 
110 & \nodata & \nodata & \nodata & 56164.7(6) & 12.9(3) & 1.4(3) & 56164.6(7) & 0.166(5) & 0.001(5) \\ 
111 & \nodata & \nodata & \nodata & 56224.2(6) & 12.0(2) & -0.2(3) & 56224.5(5) & 0.158(4) & -0.001(4) \\ 
112 & \nodata & \nodata & \nodata & 56281.6(5) & 12.4(2) & 0.6(3) & 56281.6(4) & 0.166(4) & 0.010(5) \\ 
113 & \nodata & \nodata & \nodata & 56338.5(5) & 10.9(2) & 0.6(6) & 56339.3(6) & 0.162(4) & 0.012(5) \\ 
114 & \nodata & \nodata & \nodata & 56395.2(5) & 11.2(2) & 0.7(4) & 56394.7(7) & 0.154(4) & 0.002(4) \\ 
115 & \nodata & \nodata & \nodata & 56455.1(6) & 11.0(4) & 1.0(4) & 56454.7(6) & 0.162(5) & 0.008(5) \\ 
116 & \nodata & \nodata & \nodata & 56515.6(5) & 11.0(3) & 0.5(4) & 56513.1(5) & 0.165(4) & 0.002(5) \\ 
117 & \nodata & \nodata & \nodata & 56570.7(5) & 13.1(3) & 1.5(3) & 56571.7(5) & 0.160(4) & 0.021(7) \\ 
118 & \nodata & \nodata & \nodata & 56622.8(4) & 11.8(2) & 1.4(2) & 56623.3(6) & 0.188(5) & -0.005(4) \\ 
119 & \nodata & \nodata & \nodata & 56673.3(4) & 11.4(2) & -0.3(3) & 56673.2(5) & 0.171(5) & 0.007(5) \\ 
120 & \nodata & \nodata & \nodata & 56730.6(4) & 13.3(3) & 1.8(3) & 56731.4(4) & 0.168(5) & 0.009(5) \\ 
121 & \nodata & \nodata & \nodata & 56784.4(3) & 13.1(3) & 0.1(4) & 56784.6(4) & 0.156(4) & 0.021(6) \\ 
122 & \nodata & \nodata & \nodata & 56838.8(6) & 11.5(4) & 0.4(4) & 56838.3(5) & 0.178(5) & 0.007(5) \\ 
123 & \nodata & \nodata & \nodata & 56892.7(5) & 12.5(3) & 0.8(3) & 56892.8(5) & 0.148(6) & 0.006(5) \\ 
124 & \nodata & \nodata & \nodata & 56938.2(5) & 11.4(3) & 2.2(3) & 56936.8(6) & 0.151(4) & 0.021(6) \\ 
125 & \nodata & \nodata & \nodata & 56985.3(9) & 11.6(3) & 1.8(4) & 56983.6(4) & 0.136(5) & 0.043(6) \\ 
126 & \nodata & \nodata & \nodata & 57027.5(11) & 11.3(3) & 3.5(5) & 57028.5(10) & 0.135(6) & 0.081(8) \\ 
127 & \nodata & \nodata & \nodata & 57061.9(21) & 11.4(3) & 7.2(3) & 57062.9(20) & 0.170(5) & 0.024(5) \\ 
128 & \nodata & \nodata & \nodata & 57105.3(7) & 13.6(3) & 4.3(4) & 57105.1(8) & 0.149(5) & 0.013(5) \\ 
129 & \nodata & \nodata & \nodata & 57152.8(5) & 12.9(2) & 0.4(5) & 57153.4(4) & 0.171(4) & 0.005(6) \\ 
130 & \nodata & \nodata & \nodata & 57207.0(5) & 14.3(4) & 1.2(5) & 57207.1(6) & 0.155(4) & 0.024(5) \\ 
131 & \nodata & \nodata & \nodata & 57251.5(9) & 11.1(4) & 2.6(5) & 57250.0(6) & 0.146(5) & 0.024(6) \\ 
132 & \nodata & \nodata & \nodata & 57295.1(6) & 11.8(4) & 0.2(3) & 57294.9(5) & 0.165(5) & 0.033(7) \\ 
133 & \nodata & \nodata & \nodata & 57342.8(5) & 11.3(3) & 0.7(3) & 57343.9(5) & 0.142(4) & -0.000(5) \\ 
134 & \nodata & \nodata & \nodata & 57392.7(4) & 11.6(2) & 0.3(4) & 57392.2(5) & 0.182(5) & -0.002(5) \\ 
135 & \nodata & \nodata & \nodata & 57444.8(5) & 12.6(4) & 0.8(3) & 57445.5(2) & 0.161(4) & 0.018(6) \\ 
136 & \nodata & \nodata & \nodata & 57499.7(4) & 12.4(2) & 0.9(3) & 57499.3(5) & 0.163(4) & 0.015(5) \\ 
137 & \nodata & \nodata & \nodata & 57556.4(4) & 12.3(3) & 0.2(3) & 57556.9(6) & 0.180(5) & -0.008(5) \\ 
138 & \nodata & \nodata & \nodata & 57610.2(5) & 12.8(3) & 0.4(3) & 57608.8(5) & 0.227(6) & 0.043(5) \\ 
139 & \nodata & \nodata & \nodata & 57661.0(5) & 11.4(3) & 1.4(3) & 57658.9(6) & 0.151(4) & -0.005(6) \\ 
140 & \nodata & \nodata & \nodata & 57713.1(7) & 11.6(3) & 0.9(3) & 57713.5(5) & 0.167(4) & 0.011(5) \\ 
141 & \nodata & \nodata & \nodata & 57769.9(5) & 12.6(3) & 2.0(3) & 57771.9(7) & 0.163(5) & 0.010(5) \\ 
142 & \nodata & \nodata & \nodata & 57823.1(5) & 11.5(3) & 0.2(3) & 57823.8(5) & 0.180(5) & 0.001(5) \\ 
143 & \nodata & \nodata & \nodata & 57877.9(5) & 13.8(3) & 1.1(3) & 57879.0(5) & 0.137(4) & 0.015(5) \\ 
144 & \nodata & \nodata & \nodata & 57930.9(5) & 12.5(3) & 0.5(6) & 57930.3(4) & 0.180(4) & 0.005(5) \\ 
145 & \nodata & \nodata & \nodata & 57991.1(7) & 13.0(3) & 1.4(4) & 57990.9(5) & 0.167(4) & 0.002(4) \\ 
146 & \nodata & \nodata & \nodata & 58050.2(10) & 12.0(3) & 1.8(5) & 58049.0(5) & 0.160(5) & 0.023(5) \\ 
147 & \nodata & \nodata & \nodata & 58103.3(7) & 11.7(3) & 2.0(7) & 58103.2(5) & 0.139(4) & 0.021(6) \\ 
148 & \nodata & \nodata & \nodata & 58147.5(6) & 11.6(4) & 1.9(5) & 58147.4(5) & 0.170(5) & 0.016(4) \\ 
149 & \nodata & \nodata & \nodata & 58198.1(5) & 12.4(3) & 1.2(3) & 58197.1(5) & 0.163(5) & 0.007(5) \\ 
150 & \nodata & \nodata & \nodata & 58251.0(8) & 11.6(2) & 1.4(4) & 58250.6(5) & 0.153(4) & 0.013(6) \\ 
151 & \nodata & \nodata & \nodata & 58306.4(7) & 10.5(4) & 3.3(4) & 58306.2(6) & 0.147(4) & 0.006(7) \\ 
152 & \nodata & \nodata & \nodata & 58356.3(5) & 11.5(3) & 1.7(4) & 58356.5(4) & 0.161(6) & 0.010(6) \\ 
153 & \nodata & \nodata & \nodata & 58404.6(4) & 12.4(4) & 1.7(6) & 58405.0(7) & 0.153(6) & 0.012(6) \\ 
154 & \nodata & \nodata & \nodata & 58448.5(8) & 13.6(3) & 2.7(6) & 58447.3(5) & 0.143(4) & 0.016(5) \\ 
155 & \nodata & \nodata & \nodata & 58497.6(6) & \nodata & \nodata & 58495.9(5) & \nodata & \nodata 
\enddata
\tablenotetext{a}{Arrival time (MJD) of superorbital minimum.}
\tablenotetext{b}{High-state count rate. The unit of ASM data is counts\,s$-1$, while the units of BAT and GSC data are $10^{-3}$ counts\,cm$^{-2}\,$s$-1$ and counts\,cm$^{-2}\,$s$-1$, respectively. }
\tablenotetext{c}{Low-state count rate. The unit is the same as that of the high-state count rate.}
\end{deluxetable}
\clearpage

\section{Spin Period Measurements of SMC X-1 with MAXI}\label{appendix_spin}
We list the spin period, spin-down rate, $T_0$, $T_{\pi/2}$, number of event collected with MAXI, and the corresponding $Z_2^2$ value of each superorbitl cycle in Table \ref{maxi_spin_period}.

\startlongtable
\begin{deluxetable*}{ccccccc}
\tablecaption{Spin Period Measurements of SMC X-1 with MAXI \label{maxi_spin_period}} 
\tablehead{\colhead{Cycle} & \colhead{$T_0^a$} & \colhead{$T_{\pi/2}$} & \colhead{$\nu$} & \colhead{$\dot{\nu}$} & \colhead{Events$^b$} & {$Z_2^2$}\\
\colhead{} & \colhead{(MJD)} & \colhead{(MJD)} & \colhead{(Hz)} & \colhead{($10^{-11}$ s$^{-2}$)} & \colhead{} & \colhead{}}
\startdata
91 & 55141 & 55142.9052(2) & 1.4240083(1) & 2.081(4) & 9445 & 181.8\\ 
92 & 55198 & 55197.3912(2) & 1.4241239(1) & 2.429(4) & 15692 & 267.1\\ 
93 & 55255 & 55255.7684(2) & 1.4242590(1) & 2.382(4) & 16150 & 161.7\\ 
94 & 55310 & 55310.2545(2) & 1.4243534(1) & 2.317(5) & 14334 & 170.2\\ 
95 & 55363 & 55364.7404(2) & 1.4244604(1) & 2.517(5) & 10876 & 244.2\\ 
96 & 55421 & 55419.2258(2) & 1.4245835(1) & 2.097(3) & 14014 & 168.4\\ 
97 & 55481 & 55481.4955(2) & 1.4246899(1) & 2.013(4) & 14340 & 119.0\\ 
98 & 55540 & 55539.8730(2) & 1.4248189(1) & 2.424(4) & 16324 & 158.2\\ 
99 & 55599 & 55598.2505(2) & 1.4249414(1) & 2.403(4) & 18080 & 258.3\\ 
100 & 55656 & 55656.6278(2) & 1.4250781(1) & 2.964(4) & 17376 & 182.4\\ 
101 & 55710 & 55711.1135(2) & 1.4251947(1) & 2.761(5) & 7495 & 113.3\\ 
102 & 55764 & 55765.5993(2) & 1.4253197(1) & 2.800(4) & 12980 & 91.1\\ 
103 & 55818 & 55816.1927(2) & 1.4254376(1) & 2.509(5) & 17546 & 379.0\\ 
104 & 55871 & 55870.6783(2) & 1.4255514(1) & 2.674(5) & 13521 & 188.2\\ 
105 & 55925 & 55925.1638(2) & 1.4256705(1) & 2.360(5) & 12391 & 194.4\\ 
106 & 55976 & 55975.7574(2) & 1.4257661(1) & 2.255(6) & 9755 & 142.4\\ 
107 & 56025 & 56026.3510(2) & 1.4258712(1) & 2.714(5) & 12330 & 173.2\\ 
108 & 56078 & 56076.9444(2) & 1.4259987(1) & 3.106(4) & 7902 & 74.0\\ 
109 & 56137 & 56135.3218(2) & 1.4261344(1) & 2.796(3) & 7544 & 89.0\\ 
110 & 56197 & 56197.5908(2) & 1.4262785(1) & 2.526(4) & 7573 & 68.0\\ 
111 & 56254 & 56252.0760(2) & 1.4263921(1) & 2.700(4) & 7877 & 103.4\\ 
112 & 56311 & 56310.4530(2) & 1.4265115(1) & 2.655(4) & 10384 & 98.4\\ 
113 & 56368 & 56368.8305(2) & 1.4266273(1) & 2.563(4) & 6814 & 52.8\\ 
114 & 56426 & 56427.2063(2) & 1.4267431(1) & 2.457(4) & 6643 & 51.3\\ 
115 & 56485 & 56485.5846(2) & 1.4268652(1) & 2.074(4) & 6377 & 47.5\\ 
116 & 56544 & 56543.9609(2) & 1.4269959(1) & 2.965(4) & 8626 & 86.5\\ 
117 & 56599 & 56598.4462(2) & 1.4271148(1) & 2.335(5) & 7963 & 70.3\\ 
118 & 56649 & 56649.0391(2) & 1.4272283(1) & 2.435(5) & 6212 & 87.0\\ 
119 & 56704 & 56703.5249(2) & 1.4273560(1) & 3.015(4) & 7202 & 77.2\\ 
120 & 56759 & 56758.0099(2) & 1.4274809(1) & 2.605(5) & 8359 & 137.1\\ 
121 & 56812 & 56812.4948(2) & 1.4275901(1) & 2.458(4) & 8321 & 101.8\\ 
122 & 56866 & 56866.9796(2) & 1.4276921(1) & 2.229(5) & 6260 & 58.3\\ 
123 & 56915 & 56913.6804(2) & 1.4277966(1) & 2.030(6) & 5258 & 46.2\\ 
124 & 56961 & 56960.3825(2) & 1.4278905(1) & 2.476(6) & 7269 & 67.4\\ 
125 & 57006 & 57007.0841(2) & 1.4279816(1) & 2.266(7) & 5139 & 82.3\\ 
126 & 57046 & 57046.0017(2) & 1.4280613(2) & 2.494(10) & 4491 & 82.8\\ 
127 & 57085 & 57084.9198(2) & 1.4281527(1) & 2.895(8) & 5094 & 72.1\\ 
128 & 57129 & 57127.7290(2) & 1.4282565(1) & 2.392(6) & 6905 & 67.4\\ 
129 & 57180 & 57178.3222(2) & 1.4283766(1) & 2.982(4) & 9046 & 124.1\\ 
130 & 57229 & 57228.9153(2) & 1.4284871(1) & 2.399(8) & 4918 & 57.7\\ 
131 & 57272 & 57271.7246(2) & 1.4285729(1) & 2.185(7) & 5432 & 68.5\\ 
132 & 57319 & 57318.4255(2) & 1.4286749(1) & 2.523(5) & 7664 & 38.9\\ 
133 & 57369 & 57369.0190(2) & 1.4287774(1) & 2.133(5) & 5603 & 100.6\\ 
134 & 57421 & 57419.6121(2) & 1.4288891(1) & 2.783(5) & 6790 & 70.6\\ 
135 & 57474 & 57474.0969(2) & 1.4290248(1) & 2.804(5) & 9057 & 71.4\\ 
136 & 57528 & 57528.5814(2) & 1.4291431(1) & 2.556(4) & 8719 & 80.9\\ 
137 & 57583 & 57583.0651(2) & 1.4292693(1) & 2.585(5) & 5903 & 68.2\\ 
138 & 57635 & 57633.6579(2) & 1.4293987(1) & 2.426(5) & 6992 & 37.5\\ 
139 & 57687 & 57688.1433(2) & 1.4295118(1) & 2.318(5) & 9286 & 113.9\\ 
140 & 57743 & 57742.6278(2) & 1.4296318(1) & 2.561(4) & 9221 & 82.8\\ 
141 & 57798 & 57797.1122(2) & 1.4297494(1) & 2.400(5) & 6332 & 91.1\\ 
142 & 57852 & 57851.5973(2) & 1.4298754(1) & 3.038(4) & 7903 & 74.1\\ 
143 & 57905 & 57906.0815(2) & 1.4299971(1) & 2.180(5) & 9246 & 68.5\\ 
144 & 57962 & 57960.5659(2) & 1.4301205(1) & 2.325(3) & 11013 & 130.4\\ 
145 & 58022 & 58022.8334(2) & 1.4302410(1) & 2.246(4) & 6923 & 59.6\\ 
146 & 58077 & 58077.3182(2) & 1.4303309(1) & 1.780(5) & 6978 & 54.4\\ 
147 & 58126 & 58127.9111(2) & 1.4304460(1) & 2.507(7) & 10793 & 69.1\\ 
148 & 58174 & 58174.6116(2) & 1.4305565(1) & 2.843(5) & 8053 & 69.7\\ 
149 & 58226 & 58225.2045(2) & 1.4306662(1) & 2.656(5) & 6697 & 51.2\\ 
150 & 58280 & 58279.6882(2) & 1.4307847(1) & 2.605(4) & 10275 & 79.0\\ 
151 & 58333 & 58334.1727(2) & 1.4308939(1) & 2.589(5) & 10197 & 136.6\\ 
152 & 58382 & 58380.8733(2) & 1.4310074(1) & 2.778(6) & 5782 & 55.1\\ 
153 & 58427 & 58427.5741(2) & 1.4311120(1) & 2.663(8) & 7757 & 75.7\\ 
154 & 58473 & 58474.2743(2) & 1.4312158(1) & 2.834(5) & 9173 & 77.7\\ 
155 & 58526 & 58524.8672(2) & 1.4313410(1) & 2.970(4) & 6447 & 132.1
\enddata
\tablenotetext{a}{Epoch zero of the spin frequency measurement. We set it to the midpoint of the segment.}
\tablenotetext{b}{Number of events in the energy range of 2--20\,keV. The events collected during $0.12 < \phi_{\rm{orb}} < 0.88$ and $0.1 < \phi_{\rm{sup}} < 0.6$ are used in the analysis.}
\tablecomments{This table is available online at \url{http://maxi.riken.jp/pulsar/smcx1/}}
\end{deluxetable*}

\end{document}